%% file: main.tex
\documentclass[runningheads]{llncs}
\usepackage{multirow}
\usepackage{graphicx}
\usepackage[margin=1in]{geometry}
\usepackage{xcolor}
\usepackage{xspace}
\usepackage{amssymb,amsmath,epsfig}
\usepackage{mathpartir}
\usepackage{amsthm}
\usepackage{hyperref}  
\hypersetup{
    colorlinks=true,
    linkcolor=blue, 
    urlcolor=blue,  
    citecolor=blue, 
}

\usepackage{verbatim}
\usepackage[linesnumbered, noend]{algorithm2e}
\usepackage{courier}
\usepackage{float}
\usepackage{caption}
\usepackage{mathtools}
\usepackage[capitalise]{cleveref}

\SetCommentSty{mycommfont}
\RestyleAlgo{ruled}
\input{defs}
\usepackage{orcidlink}
\begin{document}

\title{Undesignable RNA Structure Identification via Rival Structure Generation and Structure Decomposition\thanks{In Proceedings of RECOMB 2024, Lecture Notes in Computer Science (LNCS 14758), pp.~270--287, Springer.}}
\titlerunning{Undesignable RNA Structure Identification}
%
\author{Tianshuo Zhou\inst{1}\orcidlink{0009-0008-4804-0825} \and
Wei Yu Tang\inst{1}\orcidlink{0009-0008-1141-9479} \and 
David H. Mathews\inst{3,4,5}\orcidlink{0000-0002-2907-6557}  \and
Liang Huang\inst{1,2,\dagger}\orcidlink{0000-0001-6444-7045} }
\authorrunning{T.~Zhou et al.}
%

\institute{$^1$ School of EECS and $^2$ Dept.~of Biochemistry \& Biophysics, Oregon State University, Corvallis OR 97330, USA\\
$^3$ Dept.~of Biochemistry \& Biophysics, 
$^4$ Center for RNA Biology,
and $^5$ Dept.~of Biostatistics \& Computational Biology, 
University of Rochester Medical Center, Rochester, NY 14642, USA\\
$^\dagger$ \email{liang.huang.sh@gmail.com}
}
\maketitle              
\begin{abstract}
RNA design is the search for a sequence or set of sequences that will fold into predefined structures, also known as the inverse problem of RNA folding. While numerous RNA design methods have been invented to find sequences capable of folding into a target structure, little attention has been given to the identification of undesignable structures according to the minimum free energy (\MFE) criterion under the Turner model. In this paper, we address this gap by first introducing mathematical theorems outlining sufficient conditions for recognizing undesignable structures, then proposing efficient algorithms, guided by these theorems, to verify the undesignability of RNA structures. Through the application of these theorems and algorithms to the Eterna100 puzzles, we demonstrate the ability to efficiently establish that 15 of the puzzles indeed fall within the category of undesignable structures. In addition, we provide specific insights from the study of undesignability, in the hope that it will enable more understanding of RNA folding and RNA design.\\
\textbf{Availability:} Our source code is available at {\tt\url{https://github.com/shanry/RNA-Undesign}}.\\
\end{abstract}

\input{intro}
\input{rna_design}
\input{rna_undesign}
\input{algs}
\input{experiment}
\input{conclusion}

\section*{Acknowledgements}
This work was supported in part by National Institutes of Health Grant R35GM145283 (to D.H.M.) and National Science Foundation Grants 2009071 (to L.H.) and 2330737 (to L.H.~and D.H.M.). We thank the anonymous reviewers for feedbacks.

\clearpage
\bibliographystyle{splncs04}
\bibliography{references}




\clearpage

\pagenumbering{roman}

\appendix
\title{Supplementary Information}
\author{}
\institute{}
 %
%
%

%
\maketitle

\input{supp}

\end{document}

%% file: defs.tex
\newcommand{\nucA}{\ensuremath{\text{\sc A}}}
\newcommand{\nucC}{\ensuremath{\text{\sc C}}}
\newcommand{\nucG}{\ensuremath{\text{\sc G}}}
\newcommand{\nucU}{\ensuremath{\text{\sc U}}}
\newcommand{\pairs}{\ensuremath{\mathit{pairs}}\xspace}
\newcommand{\unpaired}{\ensuremath{\mathit{unpaired}}\xspace}
\newcommand{\MFE}{\ensuremath{\text{\rm MFE}}\xspace}
\newcommand{\UMFE}{\ensuremath{\text{\rm uMFE}}\xspace}

\newcommand{\LP}{\ensuremath{\mathit{loops}}\xspace}
\newcommand{\CR}{\ensuremath{\mathit{critical}}\xspace}

\newcommand{\Algb}{\ensuremath{\mathbf{Algorithm2}}}

\newcommand{\contract}{\ensuremath{\mathrm{Intersection}}}

\newcommand{\D}{\ensuremath{\mathrm{\Delta}}}
\newcommand{\DG}{\ensuremath{\mathrm{\Delta} G}}

\newcommand{\blangle}{\ensuremath{\boldsymbol{\langle}}}
\newcommand{\brangle}{\ensuremath{\boldsymbol{\rangle}}}
\newcommand{\proj}{\ensuremath{\vdash}}

\newcommand{\CMFE}{\ensuremath{\text{\rm MFE}_{\text{\rm CC}}}\xspace}
\newcommand{\CUMFE}{\ensuremath{\text{\rm uMFE}_{\text{\rm CC}}}\xspace}

\DeclareMathOperator{\defeq}{\stackrel{\Delta}{=}}
\newcommand{\undes}{\text{\texttt{undesignable}}}
\newcommand{\des}{\text{\texttt{designable}}}
\newcommand{\unknown}{\text{\texttt{unknown}}}

\renewcommand{\vec}[1]{\ensuremath{\boldsymbol{{#1}}}\xspace}

\newcommand{\vecx}{\ensuremath{\vec{x}}\xspace}
\newcommand{\vecy}{\ensuremath{\vec{y}}\xspace}
\newcommand{\vecz}{\ensuremath{\vec{z}}\xspace}

\newcommand{\ystar}{\ensuremath{\vec{y^\star}}\xspace}

\newcommand{\zstar}{\ensuremath{\vec{z^\star}}\xspace}

\newcommand{\setxi}{\ensuremath{\hat{X}}\xspace}

\newcommand{\loops}{\ensuremath{\mathit{loops}}\xspace}

\renewcommand{\angle}[1]{\ensuremath{\blangle {#1} \brangle}\xspace}

%% file: intro.tex
\section{Introduction}
Ribonucleic Acid (RNA) plays essential roles in the core activities within living cells such as transcription and translation~\cite{crick1970central}, catalyzing reactions~\cite{doudna2002chemical}, and controlling gene expression~\cite{serganov2007ribozymes}. Given a target structure, RNA design aims to find sequences that can fold into that structure. This problem, however, has been proved NP-hard~\cite{bonnet2020designing} when the simplest model of energy is adopted. The importance of RNA structure and the hardness of RNA design problem have motivated various RNA design methods~\cite{andronescu2004new,bellaousov2018accelerated,garcia2013rnaifold,hofacker1994fast,lorenz+:2011,portela2018unexpectedly,rubio2018multiobjective,taneda2011modena,Zadeh+:2010,zhou2023rna}. 

While extensive research has been dedicated to designing RNA based on a target structure, there is a notable scarcity of literature investigating the undesignability of RNA design using realistic energy models. Undesignability refers to the inability to find an RNA sequence that can fold into a desired structure using a realistic energy model. Initially, specific cases of undesignability were discovered by the work~\cite{aguirre2007computational} attempting to extend RNA-SSD~\cite{andronescu2004new}, which identified two undesignable motifs and proposed alternative motifs that would consistently be favored by the conventional Turner energy models~\cite{turner+:2010}.  Later work~\cite{halevs2015combinatorial} has presented additional motifs that prevent designability, as observed using the maximum base pair model, which is not necessarily realistic. Recent works~\cite{yao2019exponentially,yao:2021thesis} outlined a method to verify undesignability for short motifs through exhaustive enumeration and folding. To the best of our knowledge, the examination of undesignable structures or motifs based on the nearest neighbor model~\cite{Methews+:1999,Mathews+:2004,turner+:2010} has not been thoroughly explored thus far.
 

To bridge the gap between RNA design and undesignability, we propose a systematic and scientifically grounded approach known as ``Rival Structure Generation and Structure Decomposition" (RIGENDE). Our methodology operates on the principle of ``proof by construction," whereby undesignability is confirmed by the identification of rival structures that consistently outperform the target structure for any possible RNA sequence. RIGENDE can not only serve as a sanity check for empirical RNA design methods, allowing for the avoidance of executing heuristic-based algorithms in situations where no feasible solution exists, but also provide deeper insights into the energy models themselves.  This, in turn, contributes to a more profound understanding of thermodynamic models used for the prediction of RNA secondary structures. The main contributions of this paper are:
\begin{enumerate}
\item Theorems. We establish the theoretical grounds for Undesignable RNA Structure Identification, characterizing the importance of rival structure(s) and structure decomposition.
\item Algorithms. Driven by the proposed theorems, we designed and implemented highly efficient algorithms to verify undesignability automatically.
\item Application. When applying to the puzzles from Eterna100~\cite{anderson2016principles} benchmark, RIGENDE is able to prove 15 of them are undesignable. Remarkably, the verification process for each puzzle was completed within a matter of seconds or minutes.
\end{enumerate}

%% file: rna_design.tex
\section{RNA Design}
\subsection{Secondary Structure, Loop and Free Energy}

\begin{figure}[htbp]
    \begin{minipage}{.4\textwidth}
        \centering
        \includegraphics[width=\textwidth]{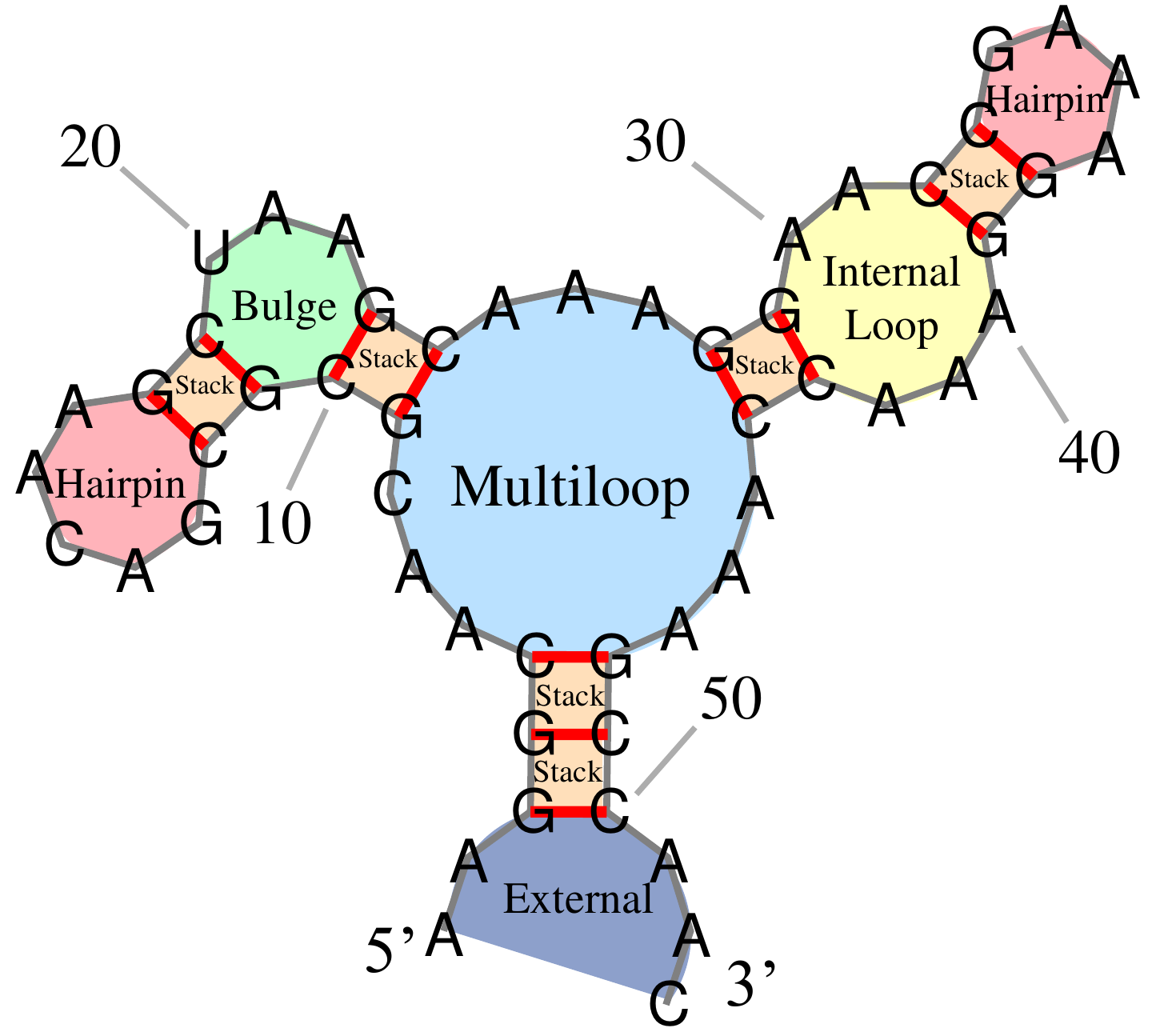}
        \captionof{figure}{An example of secondary structure and loops.} \label{fig:loop}
    \end{minipage}
    \hspace{0.07\textwidth}
    \begin{minipage}{.5\textwidth}
        \centering
        \captionof{table}{Critical positions of loops in Fig.~\ref{fig:loop}}\label{tab:cr_ex}
        \begin{tabular}{|c|c|c|}
            \hline
            \multirow{2}{*}{Loop Type}  & \multicolumn{2}{|c|}{Critical Positions} \\ \cline{2-3}
            & Closing Pairs & Mismatches (Unpaired) \\ \hline
            External & 
            (3, 50) & 2, 51 \\
            Stack & (3, 50), (4, 49) & ~ \\
            Stack & (4, 49), (5, 48) & ~ \\
            Multi & (5, 48), (9, 24), (28, 44) & 4, 49, 8, 25, 27, 45 \\
            Stack & (9, 24), (10, 23) & ~ \\
            Bulge & (10, 23), (11, 19) & ~ \\
            Stack & (11, 19), (12, 18) & ~ \\
            Hairpin & (12, 18) & 13, 17 \\
            Stack & (28, 44), (29, 43) & ~ \\
            Internal & (29, 43), (32, 39) & 30, 42, 31, 40 \\
            Stack & (32, 39), (33, 38) & ~ \\
            Hairpin & (33, 38) & 34, 37 \\ \hline
        \end{tabular}
    \end{minipage}
    
    \vspace{0.5cm}
    \begin{minipage}{1.0\textwidth}
    \centering
    \captionsetup{justification=centering} 
    \captionof{table}{Critical positions for each type of loops under Turner model implemented in ViennaRNA \\ Special hairpins~\cite{Mathews+:2004} of triloops, tetraloops and hexaloops are not considered here}\label{tab:cr}
    \begin{tabular}{|c|c|c|}
       \hline
       \multirow{2}{*}{Loop Type}  & \multicolumn{2}{|c|}{Critical Positions}\\ \cline{2-3}
       & Closing Pairs & Mismatches\\ \hline
       External & $(i_1, j_1), (i_2, j_2), \ldots, (i_k, j_k)$ & $(i_1-1, j_1+1), (i_2-1,j_2+1), \ldots, (i_k-1, j_k+1)$\\
       Hairpin & $(i, j)$ & $i+1, j-1$ \\
       Stack & $(i, j), (k, l)$ & ~ \\
       Bulge & $(i, j), (k, l)$ & ~ \\
       Internal & $(i, j), (k, l)$ & $i+1, j-1, k-1, l+1$ \\
       Multi & $(i, j), (i_1, j_1), (i_2, j_2), \ldots, (i_k, j_k)$ & $i+1, j-1, i_1-1, j_1+1, i_2-1, j_2+1, \ldots, i_k-1, j_k+1$\\ \hline
    \end{tabular}
    \end{minipage}
\end{figure}


An RNA sequence $\vecx$ of length $n$ is specified as a string of  base nucleotides $\vecx_1\vecx_2\dots \vecx_n,$  where $\vecx_i \in \{\nucA, \nucC, \nucG, \nucU\}$ for $i=1, 2,...,n$. 
A secondary structure~$\mathcal{P}$  for $\vecx$ is a set of paired indices where each pair $(i, j) \in \mathcal{P}$ indicates two distinct bases $\vecx_i \vecx_j \in \{\nucC\nucG,\nucG\nucC,\nucA\nucU,\nucU\nucA, \nucG\nucU,\nucU\nucG\}$ and each index from $1$ to $n$ can only be paired once. A secondary structure is pseudoknot-free if there are no two pairs $(i, j)\in \mathcal{P}\text{ and }(k, l)\in~\mathcal{P}$ such that  $i<k<j<l$.  In short,  a pseudoknot-free secondary structure is a properly nested set of pairings in an RNA sequence.  Alternatively, $\mathcal{P}$ can be represented as a string~$\vecy=\vecy_1\vecy_2\dots \vecy_n$,  where a pair of indices $(i, j) \in~\mathcal{P}$ corresponds to $\vecy_i=``("$, $\vecy_j=``)"$ and any unpaired index $k$ corresponds to $\vecy_k=``."$. The unpaired indices in $\vecy$ are denoted as $\unpaired(\vecy)$ and the set of paired indices in $\vecy$ is denoted as $\pairs(\vecy)$, which is equal to~$\mathcal{P}$. In nature, some RNA structures contain crossing pairings called pseudoknots. Since the computational model we use does not allow these, we do not consider them. Henceforth we elide pseudoknot-free secondary structure to just  secondary structure or structure for brevity.

The \emph{ensemble} of an RNA sequence $\vecx$  is the set of all secondary structures that  $\vecx$ can possibly fold into, denoted as $\mathcal{Y}(\vecx)$. The \emph{free energy} $\Delta G(\vecx, \vecy)$ is used to characterize the stability of $\vecy \in \mathcal{Y}(\vecx)$. The lower the free energy~$\Delta G(\vecx, \vecy)$, the more stable the secondary structure $\vecy$ for $\vecx$. In  the nearest neighbor energy model~\cite{turner+:2010},  a secondary structure is decomposed into a collection of loops, where each loop is usually a region enclosed by some base pair(s). Depending on the number of pairs on the boundary, main types of loops include hairpin loop, internal loop and multiloop, which are bounded by 1, 2 and 3 or more base pairs, respectively. In particular, the external loop is the most outside loop and is bounded by two ends ($5'$ and $3'$) and other base pair(s). Thus each loop can be identified by a set of pairs. Fig.~\ref{fig:loop}  showcases an example of secondary structure with various types of loops, where the some of the loops are notated as  

\begin{enumerate}
\item Hairpin: $H\blangle (12, 18) \brangle$.
\item Bulge: $B\blangle (10, 23), (11, 19) \brangle$.
\item Stack: $S\blangle (3, 50), (4, 49) \brangle$.
\item Internal Loop: $I\blangle (29, 43), (32, 39)\brangle$.
\item Multiloop: $M\blangle (5, 48), (9, 24), (28, 44)\brangle$.
\item External Loop: $E\blangle(3, 50) \brangle$.
\end{enumerate}
The function $\LP(\vecy)$ is used to denote the set of loops in a structure $\vecy$. The free energy of a secondary structure $\vecy$ is the sum of the free energy of each loop, 
\begin{equation}
  \Delta G(\vecx, \vecy) = \sum_{\vecz \in \loops(\vecy)} \Delta G(\vecx, \vecz),
\end{equation}
where each term $ \Delta G(\vecx, \vecz)$ is the energy for one specific loop in $\LP(\vecy)$. See Supplementary Section~\ref{sec:turner} for detailed energy functions for different types of loops in the Turner model implemented in ViennaRNA~\cite{lorenz+:2011}. The energy of each loop is typically determined by nucleotides on the  positions of enclosing pairs and their adjacent mismatch positions, which are named as \emph{critical positions} in this article. Table~\ref{tab:cr_ex} lists the critical positions for all the loops in Fig.~\ref{fig:loop} and Table \ref{tab:cr} shows the indices of critical positions for each type of loops. Additionaly, some special hairpins~\cite{Mathews+:2004} of unstable triloops and stable tetraloops and hexaloops in Turner model have a separate energy lookup table (See Supplementary Section~\ref{subsec:sp}). When evaluating the energy of a loop, it suffices to input only the nucleotides on its critical positions, i.e.,

\begin{equation}
 \Delta G(\vecx, \vecy) = \sum_{\vecz \in \loops(\vecy)} \Delta G(\vecx \proj \CR(\vecz), \vecz), \label{eq:e_sum}
\end{equation}

where  $\CR(\vecz)$ denotes the critical positions of loop $\vecz$ and  $\vecx\proj \CR(\vecz)$ denotes the nucleotides from $\vecx$ that are ``projected" onto $\CR(\vecz)$. See {Supplementary Section~\ref{sec:op}} for the detailed functionality of projection operator. The projection (\proj) allows us to focus on the relevant nucleotides for energy evaluation. For instance,
\begin{align}
  \CR(H\blangle (12, 18)\brangle) &= \{12, 13, 17, 18\},\\
  \CR(I\blangle (29, 43), (32, 39)\brangle) &= \{29, 30, 31, 32, 39, 40, 42, 43\}.
 \end{align}
For convenience of later discussion, we also interchangeably put paired positions in brackets, i.e.,  
\begin{align}
 \CR(H\blangle (12, 18)\brangle) &= \{(12, 18), 13, 17\},\\
 \CR(I\blangle (29, 43), (32, 39)\brangle) &= \{(29, 43), (32, 39) , 30, 31, 40, 42\}.
\end{align}


\subsection{MFE and Structure Distance}

The structure with the \emph{minimum free energy} is  the most stable structure in the ensemble. A structure $\ystar$ is an \MFE structure of $\vecx$, i.e. $\MFE(\vecx)$, if and only if
\begin{equation}
 ~\forall \vecy \in \mathcal{Y}(\vecx)  \text{ and }\vecy \ne \ystar , \Delta G(\vecx, \ystar) \leq  \Delta G(\vecx, \vecy). \label{eq:mfe}
\end{equation}

RNA design is  the inverse problem of RNA folding. Given a target structure $\ystar$, RNA design aims to find suitable RNA sequence $\vecx$ such that $\ystar$ is an \MFE structure of $\vecx$. For convenience, we define $\mathcal{X}(\vecy)$ as the set of all RNA sequences whose ensemble contains $\vecy$, i.e., $\mathcal{X}(\vecy)={ \vecx \mid \vecy \in \mathcal{Y}(\vecx) }$. Here we follow a more strict definition of \MFE criterion adopted in some previous studies~\cite{bonnet2020designing,halevs2015combinatorial,yao2019exponentially,ward2022fitness,zhou2023rna} on the designability of RNA, i.e., $\vecx$ is a correct design if and only if $\vecy$ is the only \MFE structure of $\vecx$, which we call unique \MFE(\UMFE) criterion to differentiate it from the traditional \MFE criterion. Formally, $\UMFE(x) = \ystar $ if and only if

\begin{equation}
 ~\forall \vecy \in \mathcal{Y}(\vecx)  \text{ and }\vecy \ne \ystar , \Delta G(\vecx, \ystar) <  \Delta G(\vecx, \vecy). \label{eq:umfe}
\end{equation}

From the perspective of optimization, the satisfaction of \MFE criterion requires that the structure distance between target structure $\ystar$ and \MFE structure of $\vecx$ is minimized to $0$. Therefore, many methods focus on optimizing  $d(\ystar, \MFE(x))$. The function $d(\vecy', \vecy'')$ represents the distance between two secondary structures $\vecy'$ and $\vecy''$, which is defined as
\begin{equation}
\begin{aligned}
d(\vecy', \vecy'') = n - 2\cdot |\pairs(\vecy')\cap \pairs(\vecy'')|  - |\unpaired(\vecy')\cap \unpaired(\vecy'')|, \label{eq:dist}
\end{aligned}
\end{equation}
where $\vecy'$ and $\vecy''$ have the same length $|\vecy'|=|\vecy''|=n$.

%% file: rna_undesign.tex
\section{Undesignability}

Based the \UMFE criterion in Eq.~\ref{eq:umfe},  the straightforward meaning of undesignability is that such a condition can not be satisfied for any RNA sequence $\vecx$ given a target structure $\ystar$. Alternatively, we give the formal definition of undesignability as follows.
\begin{definition}\label{def:umfe}
 An RNA secondary structure $\ystar$ is undesignable by \UMFE criterion if and only if
 \begin{equation}
 \forall \vecx \in \mathcal{X}(\ystar), \exists \vecy' \neq \ystar, \Delta G(\vecx, \vecy') \leq \Delta G(\vecx, \ystar).
 \end{equation}
\end{definition}
  
Similarly, we have the definition of undesignability under \MFE criterion.
  \begin{definition}\label{def:mfe}
 An RNA secondary structure $\ystar$ is undesignable by \MFE criterion if and only if
 \begin{equation}
 \forall \vecx  \in \mathcal{X}(\ystar), \exists \vecy' \neq \ystar, \Delta G(\vecx, \vecy') < \Delta G(\vecx, \ystar).
 \end{equation}
 \end{definition}
 Following previous work~\cite{halevs2015combinatorial} on undesignability, the discussions in this paper are under the setting of the \UMFE criterion and Definition~\ref{def:umfe}. However, all discussions can be straightforwardly adapted to Defition~\ref{def:mfe}.

%% file: algs.tex
\section{Theorems and Algorithms for Undesignability}\label{sec:algs}


\subsection{Algorithm 0: Exhaustive Search} %
Given a target structure $\ystar$ of length $n$, the designed sequence $\vecx$ should have the same length. Therefore, the most straightforward method is to enumerate all RNA sequences of length $n$, and check whether there exist at least one RNA sequence that can fold into $\ystar$.
Considering the designed sequence should at least satisfy that nucleotides at the paired position of the target structure should be matchable, the number of brute-force enumeration is $6^{|\pairs(\ystar)|}\times4^{|\unpaired(\ystar)|}$, as there are $6$ choices for a pair and $4$ types of nucleotides. Notice that $2 \cdot |\pairs(\ystar)|+|\unpaired(\ystar)|=n$ and the RNA folding algorithms typically have a cubic time complexity with respect to sequence length $n$, the overall complexity~$\mathcal{O}(6^{|\pairs(\ystar)|}\times4^{|\unpaired(\ystar)|}\cdot n^3)$ makes  brute-force search impractical even for very short structures. 

\subsection{Theorem 1 and Algorithm 1: Identify One Rival Structure}
One observation from RNA design is that when the designed RNA sequence $\vecx$ can not fold into the target structure $\ystar$, sometimes $\vecx$ tends to fold into another structure $\vecy'$.
Another observation is that $\vecy'$ can be very close to $\ystar$, i.e., their structure distance  $d(\ystar, \vecy')$,  can be very small. For example, when designing the puzzle ``Simple Single Bond" (shown as $\ystar$ in Fig.~\ref{fig:ex1}) from the benchmark Eterna100, the designed sequence  (shown as $\vecx$ in Fig.~\ref{fig:ex1}) tends to fold into another similar structure  (shown as $\vecy'$ in Fig.~\ref{fig:ex1}). 

\begin{figure}
\centering
  \begin{minipage}{\textwidth}  
  \centering
  \begin{tabular}{rl}
 $\ystar$&:	\verb|......(.........((((.....)))).........)......................|\\
 $\vecy'\,$&:	\verb|................((((.....))))................................|\\
 $\vecx$&:		\verb|AUAAGCGGUAAAAAAAGUGCGAAAAGCAUGAAAAAAAACAGAAAAAAAAAAAAAAAAAAAA|
 \end{tabular}
  \end{minipage}
  \caption{Example for Theorem 1 and Algorithm 1}
  \label{fig:ex1}
\end{figure}

In this instance, we know that at least for $\vecx$, $\vecy'$ is a more advantageous choice than $\ystar$. We further hypothesize that $\vecy'$ is superior to $\ystar$ for any RNA sequence that can possibly fold fold into $\ystar$. If the hypothesis holds, then we can assert that $\ystar$ is undesignable under \UMFE criterion. This leads to the first theorem we proposed.

\begin{theorem}\label{theo:1}
A structure $\ystar$ is undesignable, if
\begin{equation}
 \exists \vecy' \neq \ystar, \forall \vecx  \in \mathcal{X}(\ystar),  \Delta G(\vecx, \vecy') \leq \Delta G(\vecx, \ystar). \label{eq:theo1}
 \end{equation}
\end{theorem}

It is worth noting that the condition in Theorem~\ref{theo:1} is a special case of the condition in Definition~\ref{def:umfe}, despite both employing the same notations but in different order. The correctness of Theorem~\ref{theo:1} can be proven by the Definition~\ref{def:umfe}. Whether the undesignability can be approached by Theorem~\ref{theo:1} can be formulated as an optimization or feasibility problem.
\begin{equation}
    \label{eq:op1}
    \begin{aligned}
        {\text{find}} \quad & \vecy' \\
        \text{subject to} \quad & \Delta G(\vecx, \vecy') \leq \Delta G(\vecx, \ystar), \forall \vecx  \in \mathcal{X}(\ystar) \\
    \end{aligned}
 \end{equation}
If Theorem~\ref{theo:1} can be applied, the problem of undesignability boils down to showing that $\vecy'$ is superior to $\ystar$ for any RNA sequence, we can rewrite the inequality in Eq.~\ref{eq:theo1} as
\begin{equation}
 \Delta\Delta G(\vecy', \ystar) \defeq \Delta G(\vecx, \vecy') - \Delta G(\vecx, \ystar) \leq 0~. \label{eq:diff}
\end{equation}
Combining Eq.~\ref{eq:e_sum}, Eq.~\ref{eq:diff} can be written as the difference of two sets of energy units,
\begin{equation}
 \sum_{\vecz' \in \LP(\vecy')} \Delta G(\vecx\proj \CR(\vecz'), \vecz') - \sum_{\zstar \in \LP(\ystar)} \Delta G(\vecx \proj \CR(\zstar), \zstar) \leq 0. \label{eq:diff_z}
\end{equation}
\begin{figure}
    \begin{minipage}{.4\textwidth}
    \centering
    \includegraphics[width=\linewidth]{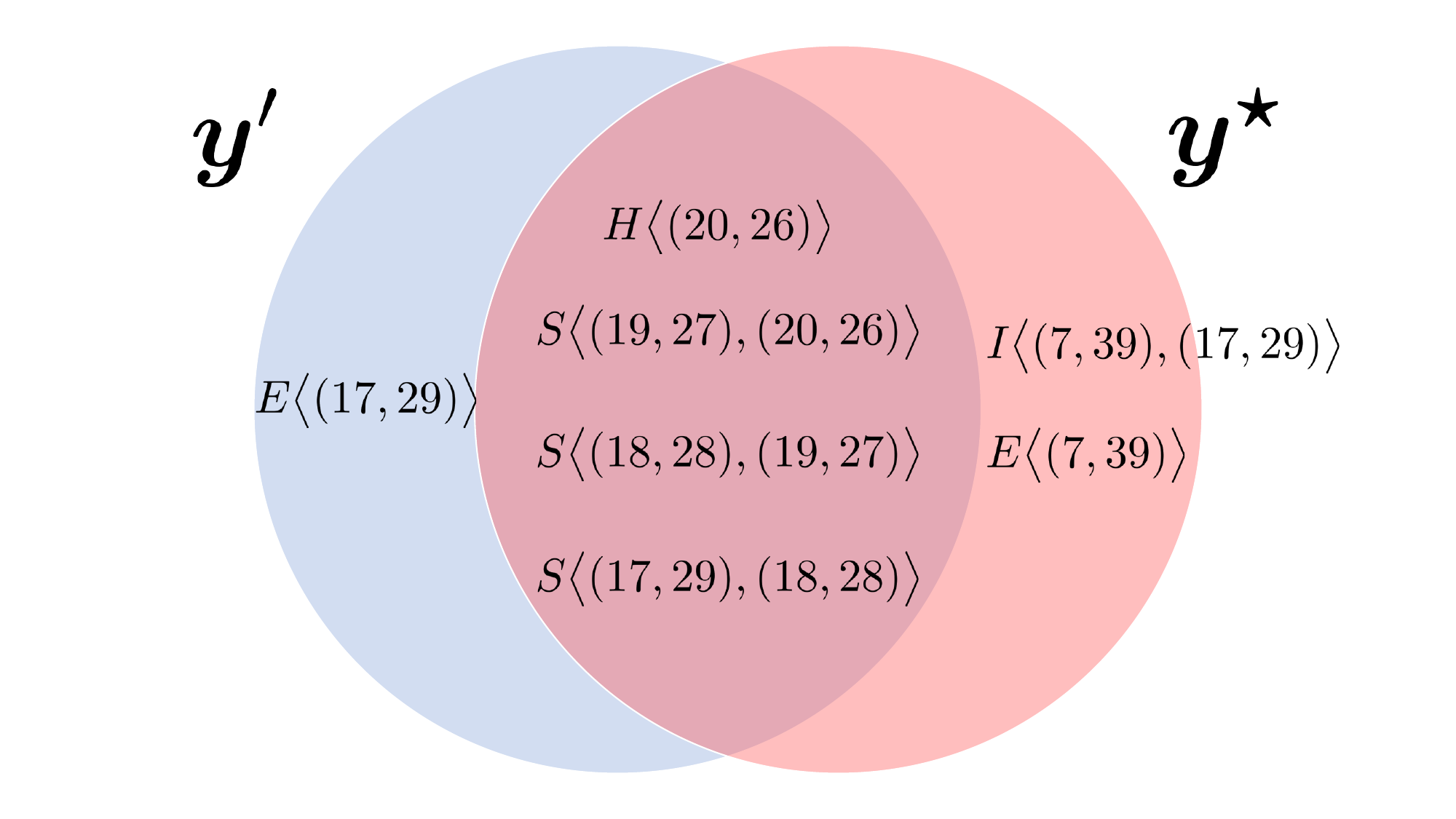}
    \captionof{figure}{Venn digram of loops in $\vecy'$ and $\ystar$.}\label{fig:venn}
  \end{minipage}
  \begin{minipage}{.5\textwidth}
    \centering
    \captionof{table}{Example of design constraint}\label{tab:x10}
    \begin{tabular}{c|cccccccccc}
	$I$ & 11 & 12 & 13  & 14  & 19 & 20 & 21 & 22\\ \hline
	$\hat{\vecx}^1$&G & G & C & G & C & C & A & U \\ \hline
	$\hat{\vecx}^2$&G & C & C & C & G & G & A & U \\ \hline
	$\hat{\vecx}^3$&G & C & U & C & G & G & A & U \\ \hline
	$\hat{\vecx}^4$&G & C & C & A & U & G & A & U \\ \hline
	$\hat{\vecx}^5$&G & C & U & A & U & G & A & U \\ \hline
	\end{tabular}
  \end{minipage}%
\end{figure}

The loops in $\LP(\vecy')$ and $\LP(\ystar)$ are compared via the Venn diagram in Fig.~\ref{fig:venn}. As we can see, the loops in $\vecy'$ and $\ystar$ overlap a lot. As a result, we can simplify Eq.~\ref{eq:diff_z} by canceling those intersected loops,
\begin{equation}
 \sum_{\!\vecz' \in \LP(\vecy')\setminus  \LP(\ystar)} \hspace{-0.8cm}\Delta G(\vecx\proj \CR(\vecz'), \vecz') - \hspace{-0.5cm}\sum_{\!\zstar \in \LP(\ystar)\setminus \LP(\vecy')} \hspace{-0.8cm}\Delta G(\vecx \proj \CR(\zstar), \zstar) \leq 0. \label{eq:diff_set}
\end{equation}

By Eq.~\ref{eq:diff_set}, the energy difference does not necessarily involve each nucleotide of $\vecx$. It is equivalent to consider only the nucleotides participating the calculation of energy difference in  Equation~\ref{eq:diff_set}, which can be written as

\begin{equation}
\begin{aligned}
 \hat{\vecx}=\vecx \proj \D(\vecy', \ystar), \Delta\Delta G(\hat{\vecx}, \vecy', \ystar) \leq 0~, \label{eq:diff_set2} \\
 \end{aligned}
\end{equation}
where
\begin{equation}
 \D(\vecy', \ystar) \defeq \bigcup_{\vecz \in \LP(\ystar)\ominus \LP(\vecy')} \CR(\vecz), \label{eq:dp}
\end{equation}
and
\begin{equation}
 \Delta\Delta G(\hat{\vecx}, \vecy', \ystar) \defeq \sum_{\vecz' \in \LP(\vecy')\setminus  \LP(\ystar)}  \hspace{-0.8cm}\Delta G(\hat{\vecx}\proj \CR(\vecz'), \vecz') - \sum_{\zstar \in \LP(\ystar)\setminus \LP(\vecy')}  \hspace{-0.8cm}\Delta G(\hat{\vecx} \proj \CR(\zstar), \zstar). \label{eq:diff_long} \\
\end{equation}
We name $ \D(\vecy', \ystar) $ as \emph{differential positions} as it is a set of all the positions whose nucleotides are involved in calculating the free energy difference between $\vecy'$ and $\ystar$.
Accordingly, each nucleotide of $\hat{\vecx}$ corresponds to one position in $ \D(\vecy', \ystar) $, and the number of nucleotides in $\hat{\vecx}$ is the same as the size of $ \D(\vecy', \ystar) $. Eq.~\ref{eq:diff_set2} implies that enumerating all possible $(\hat{\vecx}$ is equivalent to enumerating all possible $\vecx$. Suppose the number of paired positions and unpaired positions in $\D(\vecy', \ystar)$ are $p$ and $q$, the total number of enumeration would be $p^6\times q^4$. Moreover, the time cost of  evaluating Eq.~\ref{eq:diff_long} is almost $\mathcal{O}(1)$ as $\LP(\vecy')$ and $\LP(\ystar)$ only need to be computed once.
As a result,  it would be not hard to determine whether $\vecy'$ satisfies Theorem~\ref{theo:1} when $|\D(\vecy', \ystar) |$ is small, which motivates our first algorithm to efficiently verify undesignability, as described in Algorithm~\ref{alg:1}. In the case of example in Fig.~\ref{fig:ex1}, $\D(\vecy', \ystar) = \{ (7, 39), (17, 29), 6, 8,16,30, 38, 40 \}$, total number of enumerations of $\hat{\vecx}$ in Eq.~\ref{eq:diff_set2} is $147456$ which can be finished within $1$ second on a single computer in our experiments. To give a specific complexity, $\mathcal{\hat{X}}^{\D(\vecy', \ystar)}$ is used to denote the all possible nucleotide compositions at the positions in $D(\vecy', \ystar)$, we have
\begin{equation}
|\mathcal{\hat{X}}^{\D(\vecy', \ystar)}| = 6^{|\pairs({\D(\vecy', \ystar)})|} \times 4^{|\unpaired({\D(\vecy', \ystar}))|}.
\end{equation}
To prevent excessive runtime, our implementation selects only $\vec{y}'$ that is sufficiently close to $\ystar$ as input to Algorithm~\ref{alg:1}, specifically when $\mathcal{\hat{X}}^{\Delta(\vec{y}', \vec{y}^)} > M$, where $M$ is a large integer.

\SetKwInOut{KwIn}{Input}
\SetKwInOut{KwOut}{Output}
\SetKwInOut{KwPara}{Parameter}
\SetKwComment{Comment}{/* }{ */}
\begin{algorithm}[hbt!]
\caption{Identify One Rival Structure}\label{alg:1}
\KwIn{$\ystar, \vecy'$\tcp*{$\vecy' = \MFE(\vecx),~\vecx$ comes from RNA Design}}
\KwOut{$(I, \setxi)$\tcp*{$\setxi$ will store all the $\hat{\vecx}$ that violates Theorem~\ref{theo:1}}} 
$\setxi \gets \varnothing$\;
$I \gets \D(\vecy', \ystar) $\;
\ForEach {$\hat{\vecx} \in \{ \vecx \proj I \mid \vecx  \in \mathcal{X}(\ystar) \}$}
{
    \If(\tcp*[f]{If Eq.~\ref{eq:diff_set2} violated, insert $\hat{\vecx}$ into $\setxi$}){$\Delta\Delta G(\hat{\vecx}, \vecy', \ystar)  > 0$}
    {$\setxi \gets \setxi \cup \{\hat{\vecx}\}$ }
}
\Return $(I, \setxi)$\ \tcp*{If $\setxi$ is empty then $\ystar$ is undesignable}
\end{algorithm}
In fact, if $\vecy'$ is always superior to $\ystar$, any sequence $\vecx \in \mathcal{X}(\ystar)$ must be able to fold into $\vecy'$, which leads to the following corollary.  
 \begin{corollary}\label{coro:1}
 If $\vecy'$ satisfies the condition in Theorem~\ref{theo:1}, then we have $\pairs(\vecy') \subset \pairs(\ystar)$.
 \end{corollary}
\begin{proof}
Suppose there exists a pair $(i, j)$ such that $(i, j) \in \pairs(\vecy')$ but $(i, j) \notin \pairs(\ystar)$. For any sequence $\vecx$ where $\vecx_i\vecx_j$ is not among the allowed base pairs, i.e. $\vecx_i\vecx_j \notin  \{\nucC\nucG,\nucG\nucC,\nucA\nucU,\nucU\nucA, \nucG\nucU,\nucU\nucG\}$, $\vecx$ cannot fold into $\vecy'$ because $\Delta G(\vecx, \vecy') = \infty$. Therefore, if $\vecx$ prefers $\vecy'$ to $\ystar$, then $\vecy'$ cannot have any pair $(i, j)$ not in $\pairs(\ystar)$. Since $\vecy' \neq \ystar$, it follows that $\pairs(\vecy') \subset \pairs(\ystar)$.
\end{proof}
\subsection{Theorem 2 and Algorithm 2: Identify Multiple Rival Structures}
\begin{figure}[hbt!]
    \centering
    \includegraphics[width=0.6\linewidth]{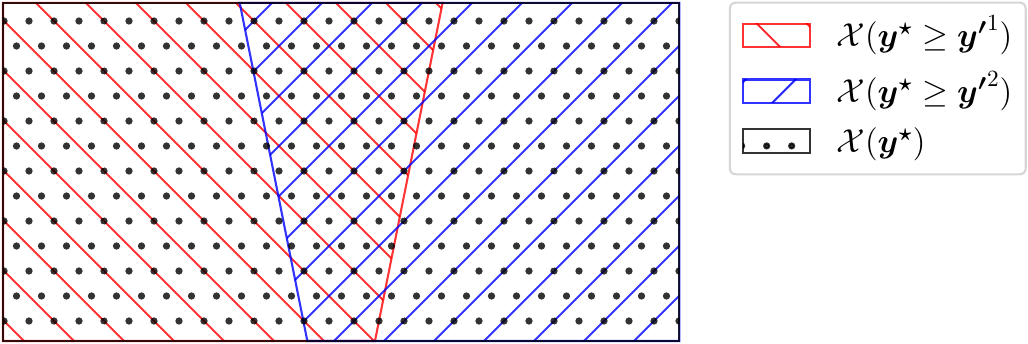}
    \captionof{figure}{Undesignability proven by 2 rival structures, $\mathcal{X}(\ystar\geq\vecy'^1) \bigcup \mathcal{X}(\ystar \geq \vecy'^2) = \mathcal{X}(\ystar)$}\label{fig:unionx}
\end{figure}
 While Algorithm~\ref{alg:1} is effective in verifying the potential rival structure $\vecy'$, it is important to acknowledge that an arbitrary structure $\vecy'$ does not necessarily always have a lower free energy than the target structure $\ystar$. Generally, the entire search space consisting of all possible RNA sequences can be divided into two subsets depending on whether each sequence $\vecx$ prefers $\ystar$ to $\vecy'$.
 \begin{proposition}\label{prop:split}
 Given a target structure $\ystar$ and another structure $\vecy' \neq \ystar$, the RNA design space $\mathcal{X}(\ystar)$ can be divided into the two sets below.
 \begin{enumerate}
 \item $\mathcal{X}(\ystar<\vecy') = \{ \vecx \mid  \Delta G(\vecx, \ystar) <  \Delta G(\vecx, \vecy'), \vecx \in \mathcal{X}(\ystar) \} $;
 \item  $\mathcal{X}(\ystar\geq\vecy') = \{ \vecx \mid  \Delta G(\vecx, \vecy') \leq \Delta G(\vecx, \ystar), \vecx \in \mathcal{X}(\ystar) \} $.
 \end{enumerate}
 \end{proposition}
 Following Proposition~\ref{prop:split}, the feasibility problem in Eq.~\ref{eq:op1} is equivalent to
 \begin{equation}
    \label{eq:op11}
    \begin{aligned}
        {\text{find}} \quad & \vecy' \\
        \text{subject to} \quad & \mathcal{X}(\ystar\geq\vecy') = \mathcal{X}(\ystar) \text{ or } \mathcal{X}(\ystar<\vecy') = \emptyset \\
    \end{aligned}
 \end{equation}
 
 In cases when no feasible $\vecy'$ satisfiying Eq.~\ref{eq:op11} can be found, we consider extending Theorem~\ref{theo:1} and feasibility problem~\ref{eq:op11} to multiple rival structures identification.
\begin{theorem}\label{theo:2}
A structure $\ystar$ is undesignable,  if 
 $$\exists Y=\{ \vecy'^1, \vecy'^2, .., \vecy'^k \} \text{ and } \ystar \notin Y, \text{such that } \forall \vecx \in \mathcal{X}(\ystar), \Delta G(\vecx, \vecy') \leq \Delta G(\vecx, \ystar)  \text{ for some } \vecy' \in Y.$$
\end{theorem}

Theorem~\ref{theo:2} can also be proven by the Definition~\ref{def:umfe}. The corresponding optimization formulation is 
\begin{equation}
    \label{eq:op2}
    \begin{aligned}
        {\text{find}} \quad & Y=\{ \vecy'^1, \vecy'^2, .., \vecy'^k \}  \\
        \text{subject to} \quad & \bigcup_{\vecy' \in Y} \mathcal{X}(\ystar\geq\vecy') = \mathcal{X}(\ystar) \text{ or } \bigcap_{\vecy' \in Y} \mathcal{X}(\ystar<\vecy') = \emptyset. \\
    \end{aligned}
 \end{equation}

Fig.~\ref{fig:unionx} shows a venn diagram when we can prove undesignability by finding a set of 2 rivial structures $Y = \{ \vecy'^1, \vecy'^2 \}$. When $\mathcal{X}(\ystar\geq\vecy'^1) \bigcup \mathcal{X}(\ystar \geq \vecy'^2) = \mathcal{X}(\ystar)$ or $\mathcal{X}(\ystar < \vecy'^1) \bigcap \mathcal{X}(\ystar < \vecy'^2) = \emptyset$, for any sequence $\vecx \in \mathcal{X}(\ystar)$, either $\vecy'^1$ or $\vecy'^2$ would have lower free energy than $\ystar$.

Given input $\ystar$ and $\vecy'$,  we call the output of  Algorithm~\ref{alg:1} a \emph{design constraint}, which characterizes the set $\mathcal{X}(\ystar<\vecy')$. For example, Fig.~\ref{fig:ex2} contains a  target structure $\ystar$ from  Eterna puzzle ``Zigzag Semicircle", along with a designed sequence $\vecx$ and $\vecy'=\MFE(\vecx)$. Upon applying Algorithm~\ref{alg:1}, the output is a tuple consisting of $ I = \D(\vecy', \ystar) = \{(11, 22), (12, 20), 13, (14, 19), 21\}$ and a set $\setxi=\{\hat{\vecx}^1, \hat{\vecx}^2, .., \hat{\vecx}^{90}\}$. Table~\ref{tab:x10} displays 5 nucleotides compositions in $\setxi$. The functionality of Algorithm~\ref{alg:1} ensures that any sequence $\vecx \in \mathcal{X}(\ystar<\vecy')$ must satisfy the design constraint $(I, \setxi)$ and any sequence satisfying  the design constraint $(I, \setxi)$ is in $\mathcal{X}(\ystar<\vecy')$. As a result, we can use $(I, \setxi)$ to represent the set $\mathcal{X}(\ystar<\vecy')$ and conduct set operations such as  intersection and union. 
\begin{figure}[h]
\centering
  \begin{minipage}{\textwidth}  
  \centering
\begin{tabular}{rl}
 $\ystar$&:	\texttt{....((((((((.(....)).).).)))))....}\\
 $\vecy'\,$&:	\texttt{....(((((((..(....)..).).)))))....}\\
 $\vecx$&:		\texttt{AAAAUGAGCCCCACGAAAGGAGAGUGCUCACAAA}
 \end{tabular}
  \end{minipage}
  \caption{Example for Theorem 2 and Algorithm 2}
  \label{fig:ex2}
\end{figure}

\begin{table}[t]
\captionsetup{justification=centering}
\caption{Example of design constraint intersection: $C'_1, C'_2 = \contract(C_1, C_2)$\\ The composition for positions $28, 29$ can only be $\nucG\nucC$ or $\nucG\nucU$}\label{tab:cc}
    \parbox[t]{.21\linewidth}{
        \centering
        \caption*{Constraint $C_1$}
        \begin{tabular}{c|cccc}
            \hline
            	$I$ & 28 & 29 & 30 & 31 \\ \hline
		$\hat{\vecx}^1$ &G  & C  & C  & C  \\ \hline
		$\hat{\vecx}^2$ &G  & U  & C  & U  \\ \hline
		$\hat{\vecx}^3$ &G  & G  & U  & U  \\ \hline
		$\hat{\vecx}^4$ &G  & G  & U  & C  \\ \hline
        \end{tabular}
    }
    \hfill
    \parbox[t]{.21\linewidth}{
        \centering
        \caption*{Constraint $C_2$}
        
       \begin{tabular}{c|cccc}
            \hline
            	$I$ & 28 & 29 & 32 & 51 \\ \hline
		$\hat{\vecx}^1$ &G  & C  & A  & G  \\ \hline
		$\hat{\vecx}^2$ &G  & C  & A  & U  \\ \hline
		$\hat{\vecx}^3$ &G  & U  & U  & C  \\ \hline
		$\hat{\vecx}^4$ &A  & G  & U  & U  \\ \hline
        \end{tabular}
     }
     \hfill
    \parbox[t]{.21\linewidth}{
        \centering
        \caption*{Constraint $C'_1$}
        
       \begin{tabular}{c|cccc}
            \hline
            	$I$ & 28 & 29 & 30 & 31 \\ \hline
		$\hat{\vecx}^1$ &G  & C  & C  & C  \\ \hline
		$\hat{\vecx}^2$ &G  & U  & C  & U  \\ \hline
        \end{tabular}
     }
\hfill
    \parbox[t]{.21\linewidth}{
        \centering
        \caption*{Constraint $C'_2$}
        
       \begin{tabular}{c|cccc}
            \hline
            	$I$ & 28 & 29 & 32 & 51 \\ \hline
		$\hat{\vecx}^1$ &G  & C  & A  & G  \\ \hline
		$\hat{\vecx}^2$ &G  & C  & A  & U  \\ \hline
		$\hat{\vecx}^3$ &G  & U  & U  & C  \\ \hline
        \end{tabular}
     }
     \vspace{-0.2cm}
\end{table}

\begin{algorithm}[hbt!]
\caption{Identify Multiple Rival Structures}\label{alg:2}
\SetKw{or}{or}
\SetKw{continue}{continue}
\SetKw{break}{break}
\SetKw{pop}{pop}
\SetKw{push}{push}
\SetKwFunction{Recurse}{\textbf{Recurse}}
\SetKwFunction{Compose}{\textbf{Compose}}
\KwIn{$\ystar, \vecx$ \tcp*{ $\vecx$ come from (unsuccessful) RNA design}}
\KwOut{ \undes/\des/\unknown} 
$\mathcal{X} \gets \mathcal{X}(\ystar)$\tcp*{Design (search) space for $\ystar$}
$Y \gets \varnothing$\tcp*{Contains all potential rival structures $\vecy'$}
$Q \gets \{ \vecx \}$\tcp*{A queue contains RNA sequences for folding}
\While{$Q$ is not empty}{
      $\vecx \gets \pop(Q)$\;
       \If{$\UMFE(\vecx) = \ystar$}{
    	\Return \des \tcp*{Identify designable case} \label{line:des}}
    $\vecy' = \MFE(\vecx)$\;
     \lIf(\tcp*[f]{Stop if no new $\vecy'$}){$ \vecy' \in Y$}{
		\break
	}
    \If(\tcp*[f]{Continue if too many enumeration in Algorithm~\ref{alg:1} }){$| \mathcal{\hat{X}}^{\D(\vecy', \ystar)}| > M$}{
		\continue
    }
    $\mathcal{X}(\ystar<\vecy') \gets \mathbf{Algorithm~\ref{alg:1}}(\ystar, \vecy') $\;
    $\mathcal{X} \gets \mathcal{X} \cap \mathcal{X}(\ystar<\vecy')$ \tcp*{Set intersection}\label{line:inter}
    $Y \gets Y \cup \{ \vecy' \}$\;
      \lIf{$\mathcal{X} = \varnothing$}{
		\Return \undes
	}
	\lIf(\tcp*[f]{Stop if Y is too large}){$ |Y| > N$}{
		\break
	}
    \If{Q is empty}{
    \For(\tcp*[f]{Sample at most $K$ sequences}){$i=1$ \KwTo $K$}{
    Sample $\vecx_{new} \in \mathcal{X}$\;
    $Q \gets \push(Q, \vecx_{new}$)\;
}
}
}
\Return \unknown
\end{algorithm}

A high level algorithm for solving Eq.~\ref{eq:op2} is shown in Algorithm~\ref{alg:2}.
Starting from a seed $\vecx$ and $\vecy'=\MFE(\vecx)$,  Algrithm~\ref{alg:2} repeatedly calls Algorithm~\ref{alg:1} to get new potential $\vecy'$ and corresponding design constraint $\mathcal{X}(\ystar<\vecy')$. Each new design constraint is intersected (line~\ref{line:inter}) with all the other design constraints previously found.  An example of design constraint intersection is show in Table~\ref{tab:cc}. Refer to {Supplementary Section~\ref{sec:op}} for specific steps of constraint intersection. The algorithm stops if no new rival structure candidate can be found or there are too many rival structures in $Y$. Algorithm~\ref{alg:2} has 3 parameters: (1) $M$ is the maximum enumeration allowed for Algorithm~\ref{alg:1}; (2) $N$ is maximum number of rival structures allowed in $Y$; (3) $K$ is the number of sampled sequences from design space $\mathcal{X}$.  At most $N$ set intersections are executed, and at most $NK$ sequences are sampled and folded. Assuming hash sets are used, the time complexity of set intersection is $\mathcal{O}(M)$. Therefore, the overall complexity of Algorithm~\ref{alg:2} is $\mathcal{O}(NM+NKn^3)$, where $n$ is the length of the input target structure.

\subsection{Theorem 3 and Algorithm 3: Structure Decomposition}
While Algorithm~\ref{alg:1} and~\ref{alg:2} are  efficient when the input $\D(\ystar, \MFE(\vecx))$ is small, it is not practical otherwise. For instance, Fig.~\ref{fig:ex3} showcases the puzzle ``multilooping fun" from Eterna100 benchmark. The difference between $\ystar$ and $\vecy'$ is so huge that is not suitable as input for Algorithm~\ref{alg:1} and~\ref{alg:2}. It is worth noting that a base pair $(i, j) \in \ystar$ divides the free energy $\Delta G(\vecx, \ystar)$ into two uncoupled parts: one within and one outside $\ystar_{i\rightarrow j} = \ystar_i\ystar_{i+1}\ldots \ystar_{j}$, respectively,
\begin{equation}
  \Delta G(\vecx, \ystar) = \sum_{\vecz \in \loops(\ystar), \vecz \in \loops(\ystar_{i\rightarrow j})} \Delta G(\vecx, \vecz) + \sum_{\vecz \in \loops(\ystar), \vecz \notin \loops(\ystar_{i\rightarrow j})} \Delta G(\vecx, \vecz).
\end{equation}
 When it is impractical to apply Algorithm~\ref{alg:1} and~\ref{alg:2} to the entire structure $\ystar$, it might be beneficial to search for rival structures for a pair-bounded \emph{substructure} $\ystar_{i\rightarrow j}$. For example, if another pair-bounded $\vecy''_{i\rightarrow j}$ is always more advantageous, then replacing $\ystar_{i\rightarrow j}$ with $\vecy''_{i\rightarrow j}$ in $\ystar$ will yield another structure $\vecy''$ that qualifies as a rival structure for $\ystar$. However, the crucial point for such a decomposition and combination is that both $\ystar_{i\rightarrow j}$ and $\vecy''_{i\rightarrow j}$ must be enclosed by a pair, ensuring that the free energy is the sum of the energy of loops within the pair and outside the pair $(i, j)$. Therefore, we propose to decompose a target structure by base pairs such that the undesignability of a pair-bounded substructure can assure the undesignability of the original target structure.
\begin{figure}[h]
\centering
  \begin{minipage}{\textwidth}  
  \centering
  \begin{tabular}{rl}
 $\ystar$: &	\verb|((.(..(.(....).(....).)..).(....).))|\\[-0.2cm]
            &  \verb|      -----------------           |\\[-0.2cm]
 $\vecy'\,$: &	\verb|((............((....))............))|\\
 $\quad \ystar_{i\rightarrow j}~\;$:	&\verb|      (.(....).(....).)        |$\quad\quad\quad\quad\quad$\\
 $\quad \vecy''_{i\rightarrow j}\;\;$:	&\verb|      (...............)          |$\quad\quad\quad\quad\quad$\\
 $\vecy''\quad$: &	\verb|((.(..(...............)..).(....).))|\\[-0.2cm]
             &  \verb|      -----------------           |\\[-0.2cm]
 \end{tabular}
  \end{minipage}
  \caption{Example for Theorem 3 and Algorithm 3}
  \label{fig:ex3}
\end{figure}

\begin{definition}
A structure $\vecy=\vecy_1\vecy_2\ldots \vecy_n$ is context-constrained if $(1, n)\in \pairs(\vecy)$, i.e., its first and last positions are paired.
\end{definition}
 For a sequence $\vecx=\vecx_1\vecx_2\ldots \vecx_n$ satisfying $\vecx_1 \vecx_n \in \{\nucC\nucG,\nucG\nucC,\nucA\nucU,\nucU\nucA, \nucG\nucU,\nucU\nucG\}$,
its \emph{context-constrained ensemble} is defined as $\mathcal{Y}_{CC}(\vecx) = \{ \vecy \mid (1, n)\in \pairs(\vecy), \vecy \in \mathcal{Y} (\vecx) \}$.
A context-constrained structure $\ystar$ is a context-constrained \MFE structure of $\vecx$, i.e., $ \CMFE(\vecx)$, if and only if
\begin{equation}
 \forall \vecy \in \mathcal{Y}_{CC}(\vecx) \text{ and }\vecy \neq \ystar , \text{~then~} \Delta G(\vecx, \ystar) \leq \Delta G(\vecx, \vecy). \label{eq:cmfe}
\end{equation}
A context-constrained structure $\ystar$ is the context-constrained \UMFE structure of $\vecx$, i.e., $ \CUMFE(\vecx)$, if and only if
\begin{equation}
 \forall \vecy \in \mathcal{Y}_{CC}(\vecx) \text{ and }\vecy \neq \ystar , \text{~then~} \Delta G(\vecx, \ystar) <  \Delta G(\vecx, \vecy). \label{eq:cumfe}
\end{equation}
 Accordingly, we can define context-constrained undesignability by \UMFE criterion.
 \begin{definition}
A context-constrained structure $\ystar$ is context-constrained-undesignable if and only if
$$\forall \vecx \in \mathcal{X}{(\ystar)}, \exists \vecy' \neq \ystar \text{~and $\vecy'$ is context-constrained}, \Delta G(\vecx, \vecy') \leq \Delta G(\vecx, \ystar). $$ \label{def:ccud}
\end{definition}
 The above definitions allow us to succinctly express the idea of proving undesignability via structure decomposition in Theorem~\ref{theo:3}. 

\begin{algorithm}[h!]
\caption{Identify Rival Structures with Structure Decomposition}\label{alg:3}
\SetKw{or}{or}
\SetKw{and}{and}
\SetKw{continue}{continue}
\SetKwFunction{Recurse}{Recurse}
\KwIn{$\ystar, \vecx$ \tcp*{$\vecx$ comes from RNA Design}}
\KwOut{\undes/\unknown}
 \ForEach {$(i, j)\in \pairs(\ystar)$}{
 	\If(\tcp*[f]{Hairpin\&stack excluded}){$H\blangle (i, j) \brangle \notin \LP(\ystar)~\and~S\blangle (i, j), (i+1, j-1) \brangle \notin \LP(\ystar)$}{
		\If(\tcp*[f]{Constrained folding}){$\ystar_{i\rightarrow j} \neq \CUMFE(\vecx_{i\rightarrow j})$}{  
			\If(\tcp*[f]{Use $\CMFE\&\CUMFE$ in Alg.\ref{alg:2}}){$\Algb(\ystar_{i\rightarrow j}, \vecx_{i\rightarrow j}) =$ {\rm \undes}}{
			\Return \undes
			}
		}
	}
}
\Return \unknown\;
\end{algorithm}
\begin{theorem}
A structure $\ystar$ is undesignable if there exists a pair $(i, j) \in \pairs(\ystar) $ such that the structure $\ystar_{i\rightarrow j}$ is context-constrained undesignable, where $\ystar_{i\rightarrow j}=\ystar_i\ystar_{i+1}\ldots \ystar_{j}$. \label{theo:3}
\end{theorem}
\begin{proof}
 By Def.~\ref{def:ccud}, $\forall \vecx_{i\rightarrow j} \in \mathcal{X}{(\ystar_{i\rightarrow j})}, \exists \vecy'_{i\rightarrow j} \neq \ystar_{i\rightarrow j} \text{~and $\vecy'_{i\rightarrow j}$ is context-constrained}, \Delta G(\vecx_{i\rightarrow j}, \vecy'_{i\rightarrow j}) \leq \Delta G(\vecx_{i\rightarrow j}, \ystar_{i\rightarrow j})$. We can construct a structure $\vecy'' \neq \ystar$ by substituting $\ystar_{i\rightarrow j}$ within $\ystar$ with $\vecy'_{i\rightarrow j}$ such that $\loops(\vecy'')=\loops(\ystar)\setminus\loops(\ystar_{i\rightarrow j})\cup\loops(\vecy'_{i\rightarrow j})$. As a result, $ \forall \vecx \in \mathcal{X}{(\ystar)}, \exists \vecy'' \neq \vecy, \Delta G(\vecx, \vecy'') - \Delta G(\vecx, \ystar)=  \Delta G(\vecx_{i\rightarrow j}, \vecy'_{i\rightarrow j}) -  \Delta G(\vecx_{i\rightarrow j}, \ystar_{i\rightarrow j})) \leq 0$.

 \end{proof}
The algorithm for Theorem~\ref{theo:3} is presented in Algorithm~\ref{alg:3}. Each substructure of the target structure bounded by a pair $(i, j)$ can be regarded as a context-constrained structure, which is input to Algorithm~\ref{alg:2}. For efficiency we excluded those pairs enclosing a hairpin loop or a stack loop. The target structure is then proven undesignable if one decomposed substructure is verified to be context-constrained undesignable. Since Algorithm~\ref{alg:2} is called at most $n/2$ times, the overall complexity of Algorithm~\ref{alg:3} is $\mathcal{O}(NMn+NKn^4)$.

%% file: experiment.tex

\section{Experiments on Eterna100 Dataset}
\subsection{Setting}
We applied the three algorithms described in Section~\ref{sec:algs} to structures from the Eterna100 dataset~\cite{anderson2016principles}, a well-known benchmark for RNA inverse folding. Eterna100 contains a list of 100 secondary structure design challenges (also called puzzles) with a wide range of difficulties.

A previous study~\cite{koodli2021redesigning} identified 19 puzzles were never successfully designed with the  folding parameters of ViennaRNA 2.5.1. We took a step further and tried to prove that it is impossible to solve some of Eterna100 puzzles under the \UMFE criterion.

For each target structure $\ystar$ in Eterna100, we first attempted RNA design using two state-of-the-art methods NEMO~\cite{portela2018unexpectedly} and SAMFEO~\cite{zhou2023rna}. We chose the two because our previous studies show that they were able to solve the most puzzles~\cite{zhou2023rna}. We adopted the same setting as the RNA design experiments in SAMFEO~\cite{zhou2023rna}. Eventually, we obtained 22 structures that neither of the two programs designed successfully under the \UMFE criterion with ViennaRNA 2.5.1 parameters. For each unsolved puzzle $\ystar$, we selected the output $\vecx$ such that its \MFE structure $\vecy'$ has the minimal structure  distance $d(\ystar, \vecy')$,  then we used $\ystar$, $\vecy'$, and $\vecx$  as the input to our algorithms.

The three algorithms are implemented in \verb|C++| and running on Linux, with 3.40 GHz Intel Xeon E3-1231 CPU and 32G memory. Our implementation also utilized OpenMP to achieve parallelization and the program was ran with 8 CPUs. In Algorithm \ref{alg:2}, we used LinearFold~\cite{huang+:2019} (beam size set as $0$, which means exact search without beam pruning) with the energy parameter from ViennaRNA 2.5.1 to find $\MFE(\vecx)$ and $\UMFE(\vecx)$. LinearFold also provides the functionality for constraint folding, which is used in Algorithm \ref{alg:3} to obtain $\CMFE(\vecx_{i\rightarrow j})$ and $\CUMFE(\vecx_{i\rightarrow j})$. Notice our algorithms do not  rely on any specific folding package, our released implementation also support using ViennaRAN package for folding and constrained folding which will yield the same output. To prevent the algorithms from running indefinitely, we set the parameters in Algorithm 2 as follows: $M=10^{10},~N=10^5,~K=500$.
\subsection{Results}
The results of applying our algorithms are presented in Table \ref{table:results}. In total, we identified that 15 out of those the 22 puzzles are undesignable using ViennaRNA 2.5.1 parameters. The algorithms identified the rival structure(s) for 14 out of those 15 undesignable puzzles automatically. For the Puzzle 87, the algorithms took the puzzle and a candidate rival (sub)structure we manually selected then proved the puzzle is undesignable according to Theorem~\ref{theo:3}. 

The implementation of our algorithm also enable turning off special hairpins in energy model. As a result, in addition to the aforementioned 15 undesignable puzzles, our algorithms can automatically prove the Puzzle 50 is undesignable when special hairpins are not considered.

An additional noteworthy finding is that Algorithm~\ref{alg:2} can also identify a \UMFE (or \MFE) solution (line~\ref{line:des}) in the process of searching rival structure candidates by folding new sequences. Remarkably, the puzzle ``Short String 4" (in the 22 unsolved puzzles) turned out to be designable, i.e., Algorithm~\ref{alg:2} successfully generated an RNA sequence that adopts the target structure as the unique \MFE structure. See {Supplementary Section~\ref{sec:dcase}} for the structure of  ``Short String 4" along with the designed sequence. Finally, the designability of remaining $5$ puzzles remains uncertain, whose puzzles names are \texttt{Taraxacum officinale, Mat-Lot2-2B, Gladius, Hoglafractal}, and \texttt{Teslagon}. 

\subsection{Insights}
For those puzzles proven undesignable by our three algorithms, we further compared the identified rival structures\footnote{We released those rival structures at \texttt{\url{https://github.com/shanry/RNA-Undesign/tree/main/data/results/rigend}}.} with original target structures.  Main insights are summarized as follows.
\begin{enumerate}
	\item The undesignability identified by Algorithm~\ref{alg:1}  is usually caused by some lonely pair or double pairs in the target structure, which is consistent to the observation of previous study~\cite{anderson2016principles} on the difficulty of puzzles. For example, the $\vecy'$ in Fig.~\ref{fig:ex1} has one less pair compared to $\ystar$. However, our approach can provide loop-level reasoning and quantitative explanation, which goes beyond heuristics.
	\item However, contrary to the principle~\cite{anderson2016principles} that symmetry is a feature of difficulty for RNA design, we found that the undesignability is usually caused by some independent local region in a target structures, as is highlighted in the structures plot in Table~\ref{table:results}.
	\item If an undesignable structure can be proven by identifying multiple rival structures, the number of the rivals tends to be small and those rival structures can be very similar to each other. There are $4$ cases with multiple rival structures in Table~\ref{table:results}, and their number of rivial structures are $8, 9, 9, 2$ respectively.
	\item The constrained-context undesignable structures identified by Algorithm~\ref{alg:3} often contains some hairpin enclosed by a single pair or double pairs as the cases highlighted in Table~\ref{table:results}. However, it is hard to locate those regions by attempting RNA design and find a $\vecy'$ similar to target structure $\ystar$, which demonstrates the cruciality of structure decomposition.
\end{enumerate}
\renewcommand{\thempfootnote}{\arabic{mpfootnote}}
\begin{table}[!h]
    \caption{List of Eterna100 puzzles that we prove to be undesignable, with context-constrained-undesignable substructures highlighted. If \#Rivals is $1$, the rival structure can be obtained by removing the red-colored pair(s) from the target structure.}
    \begin{minipage}{\textwidth}
    \centering
    \begin{tabular}{|l| l| r| r| r| r| c| }
    \hline
        Id & Puzzle & \scalebox{0.8}{Length} &\scalebox{0.8}{\#Rivals}& Algorithm & \scalebox{0.8}{Time (sec.)} & Structure  \\ \hline
        50 & 1, 2, 3 and 4 bulges\footnote[6]{This puzzle is proven undesignable if we ignore energies of special hairpins.} & 105 & 1 & alg 1 & 0.08 & \raisebox{-.5\height}{\includegraphics[width=0.09\textwidth]{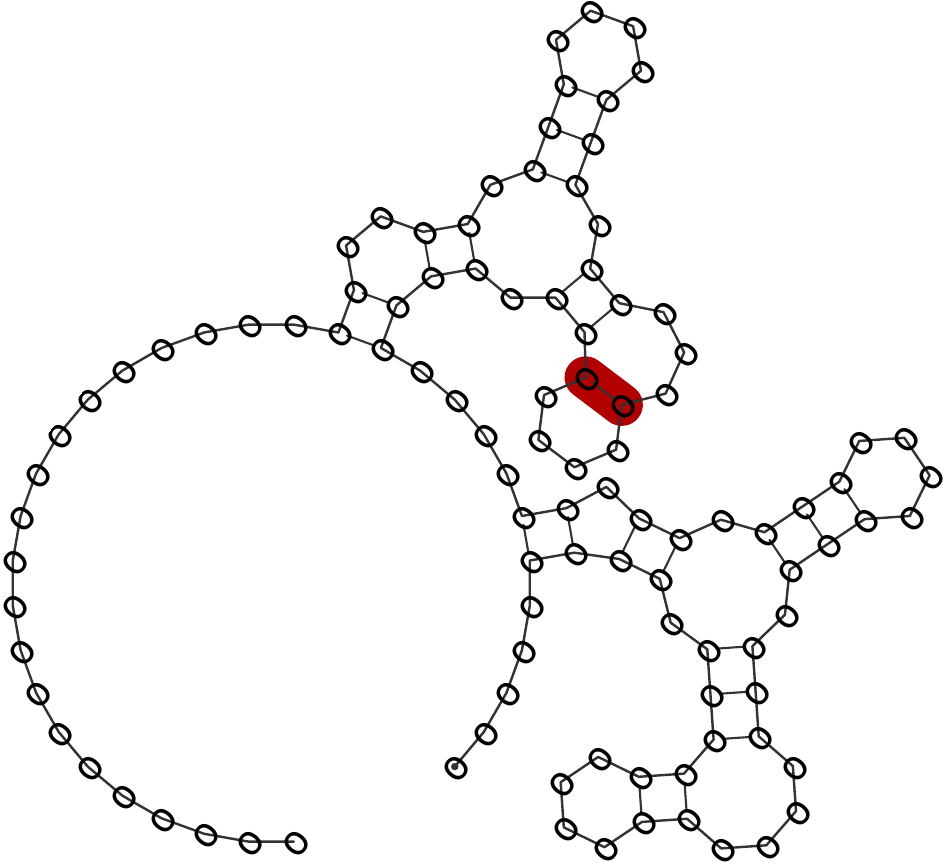}} \\ \hline
        52 & \scalebox{0.8}{[RNA] Repetitious Seqs.~8/10} & 80 & 1 & alg 1 & 0.03 & \raisebox{-.5\height}{\includegraphics[width=0.09\textwidth, angle=0]{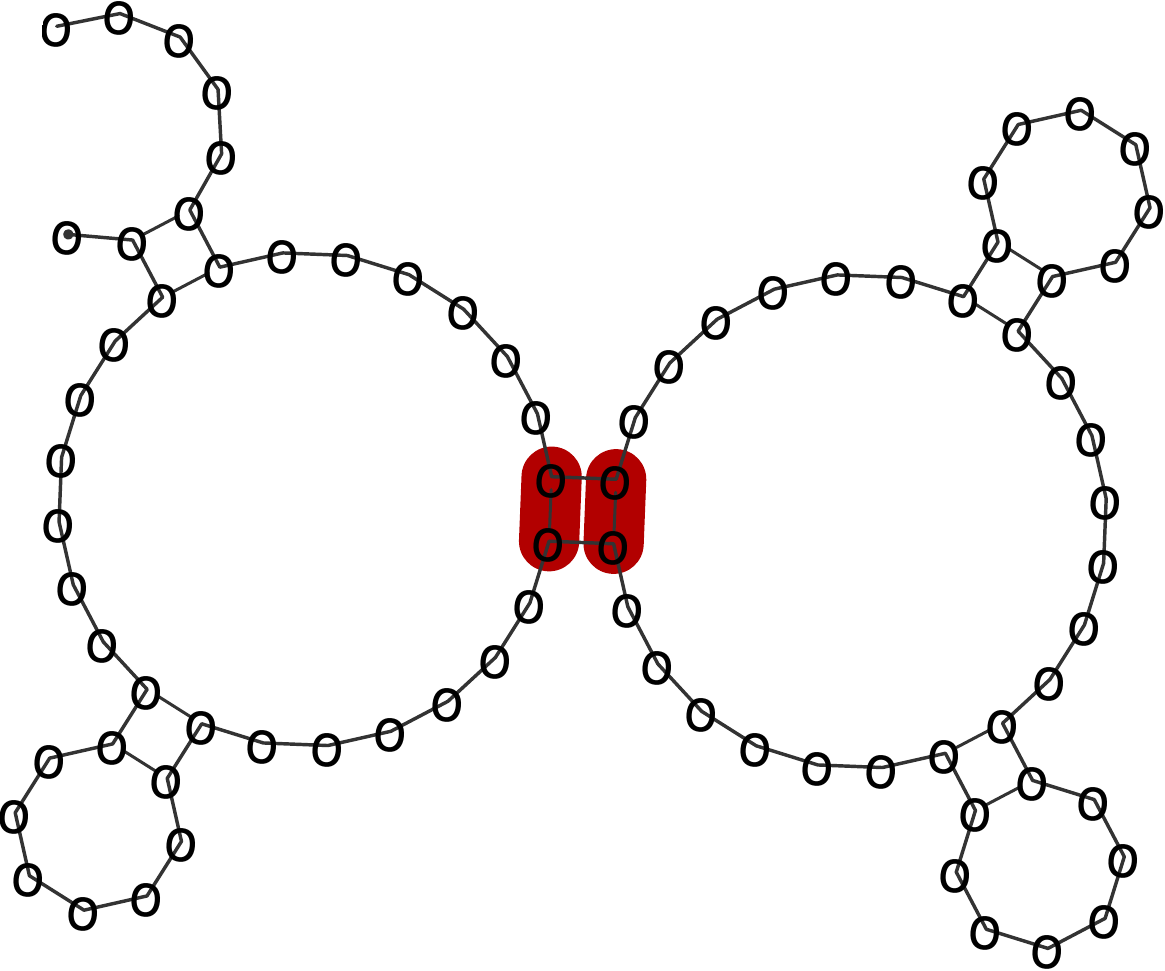}}  \\ \hline
        57 & multilooping fun & 36 & 1 & alg 3 & 22.74 & \raisebox{-.5\height}{\includegraphics[width=0.08\textwidth]{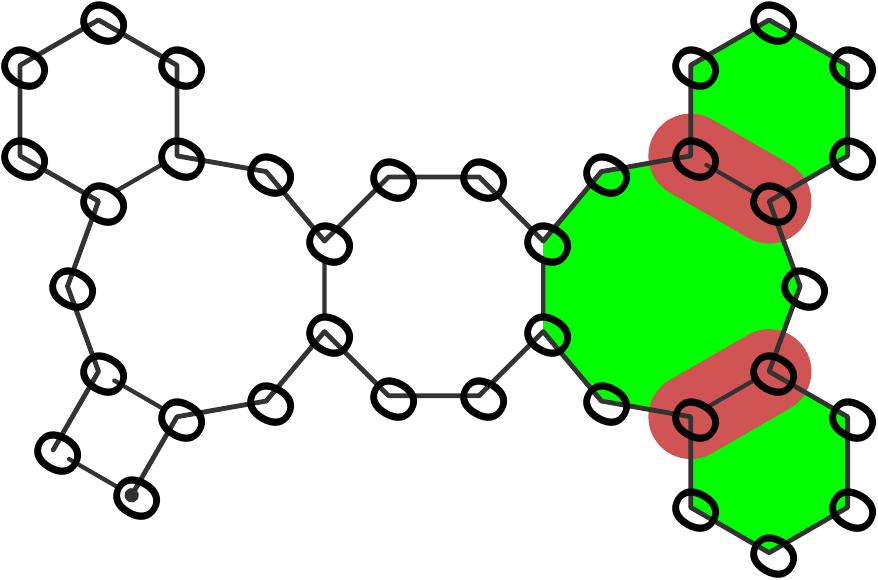}} \\  \hline
        60 & \scalebox{0.8}{Mat - Elements \& Sections} & 105 & 8 & alg 2 & 1.82 & \raisebox{-.5\height}{\includegraphics[width=0.18\textwidth, angle=0]{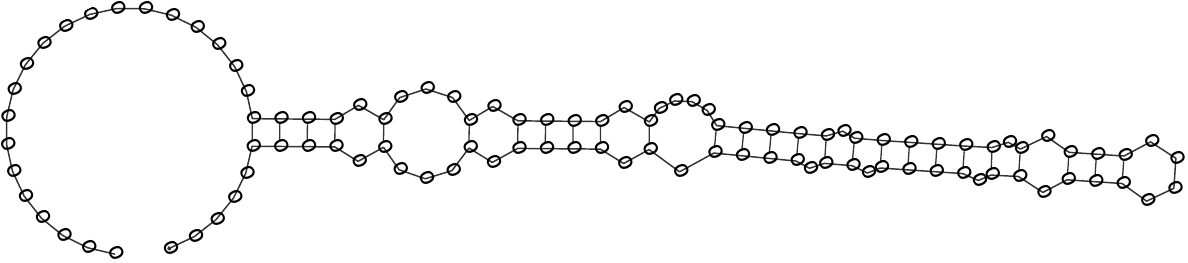}}  \\  \hline
        61 & Chicken feet & 67 & 1 & alg 3 & 231.61 & \raisebox{-.5\height}{\includegraphics[width=0.12\textwidth]{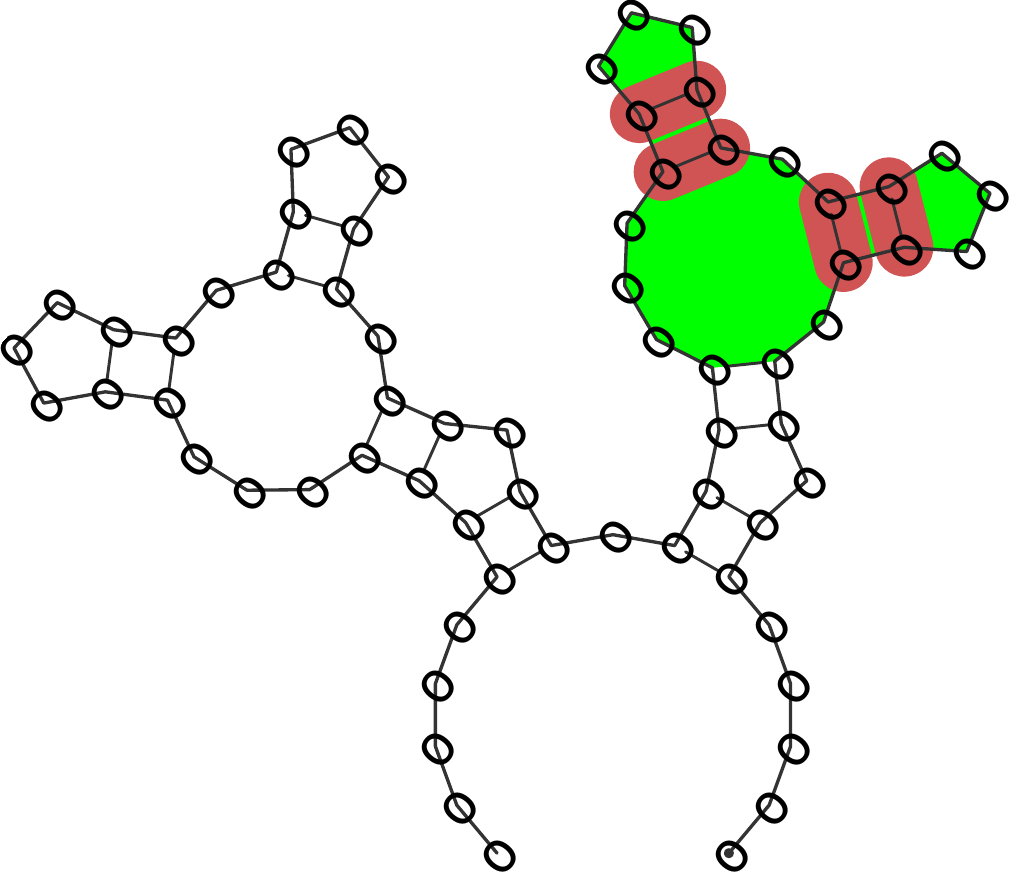}}   \\  \hline
        67 & Simple Single Bond & 61 & 1 & alg 1 & 0.10 &\raisebox{-.5\height}{\includegraphics[width=0.12\textwidth, angle=0]{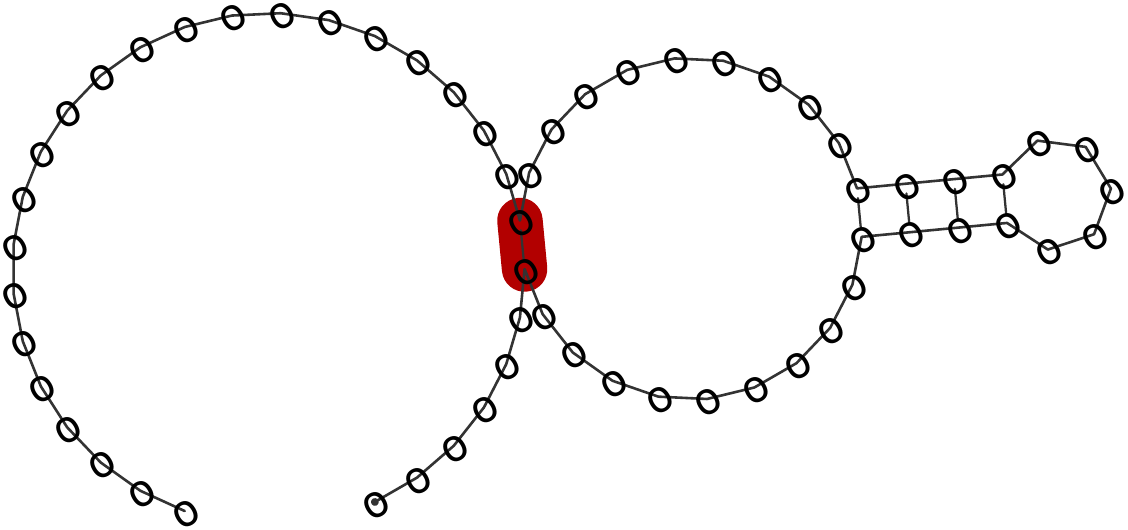}}   \\  \hline
        72 & Loop next to a Multiloop\footnote[7]{Though this puzzle is proven undesignable by the \UMFE criterion, it is designable by the \MFE criterion.} & 73 & 1 & alg 1 & 0.19 &\raisebox{-.5\height}{\includegraphics[width=0.11\textwidth, angle=0]{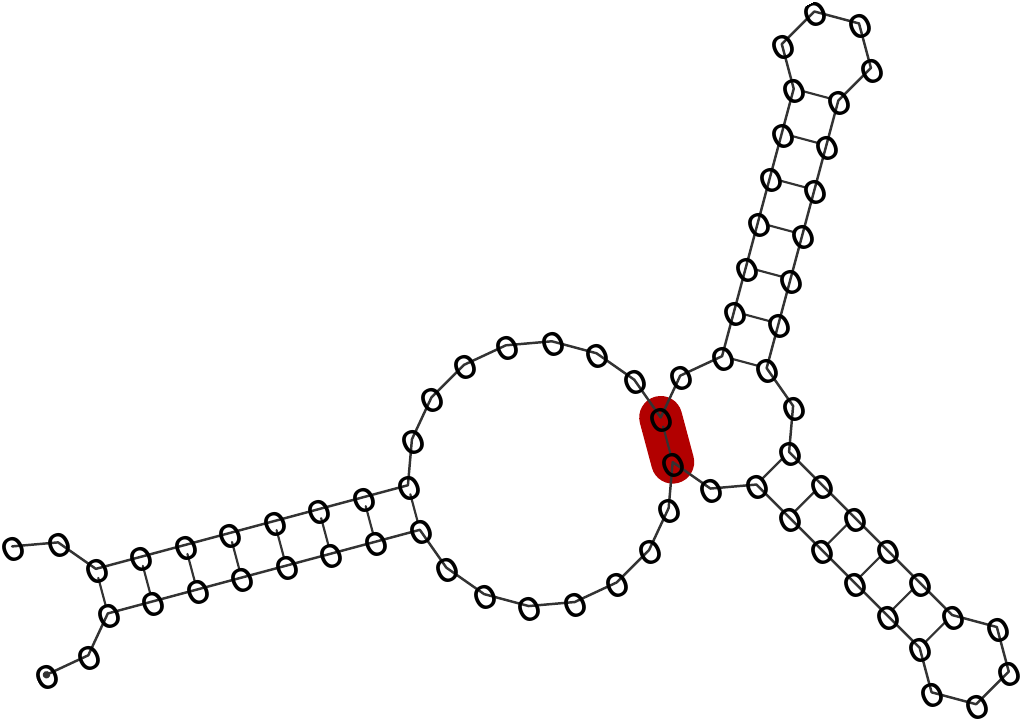}}   \\  \hline
        80 & Spiral of 5's & 397 & 1 & alg 1 & 0.17 &\raisebox{-.5\height}{\includegraphics[width=0.09\textwidth, angle=0]{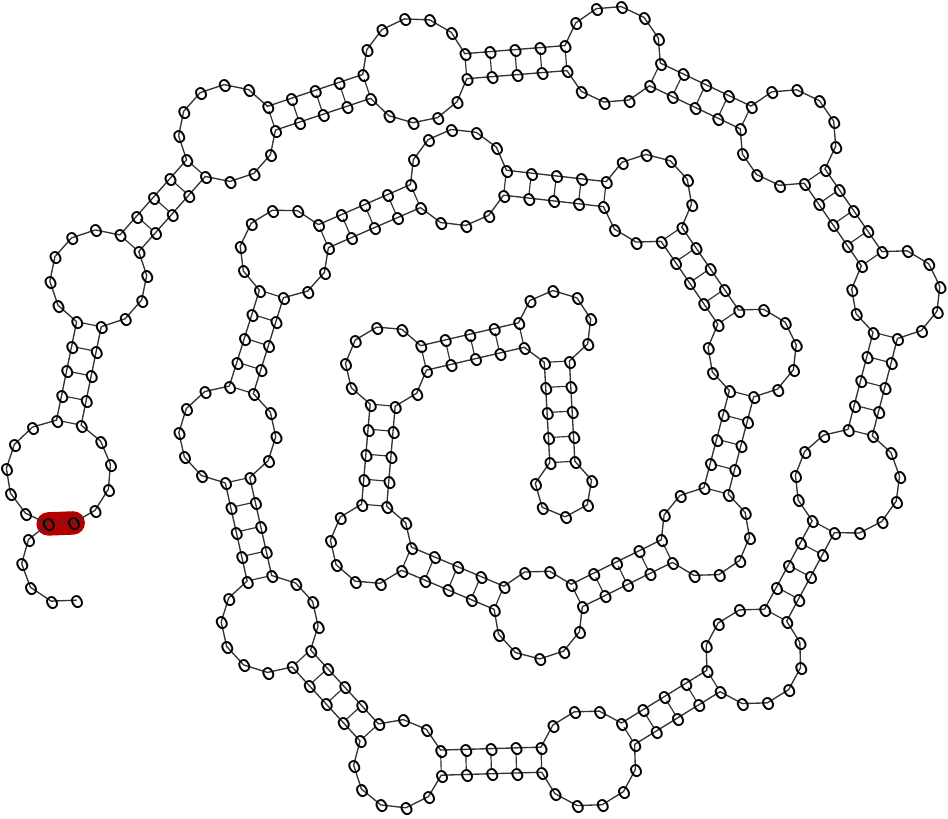}}   \\  \hline
        81 & Campfire & 212 & 9 & alg 3 & 0.25  &\raisebox{-.5\height}{\includegraphics[width=0.09\textwidth, angle=0]{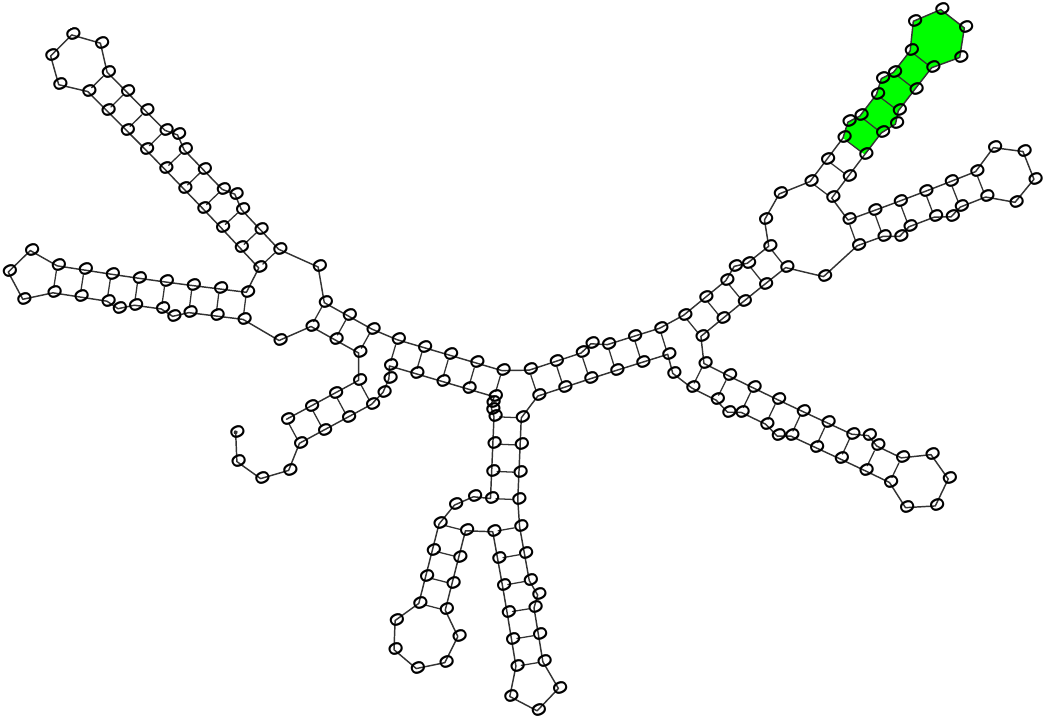}}    \\  \hline
        86 & \scalebox{0.8}{Methaqualone C$_{16}$H$_{14}$N$_2$O} & 355 & 1 & alg 3 & 17.66 &\raisebox{-.5\height}{\includegraphics[width=0.13\textwidth]{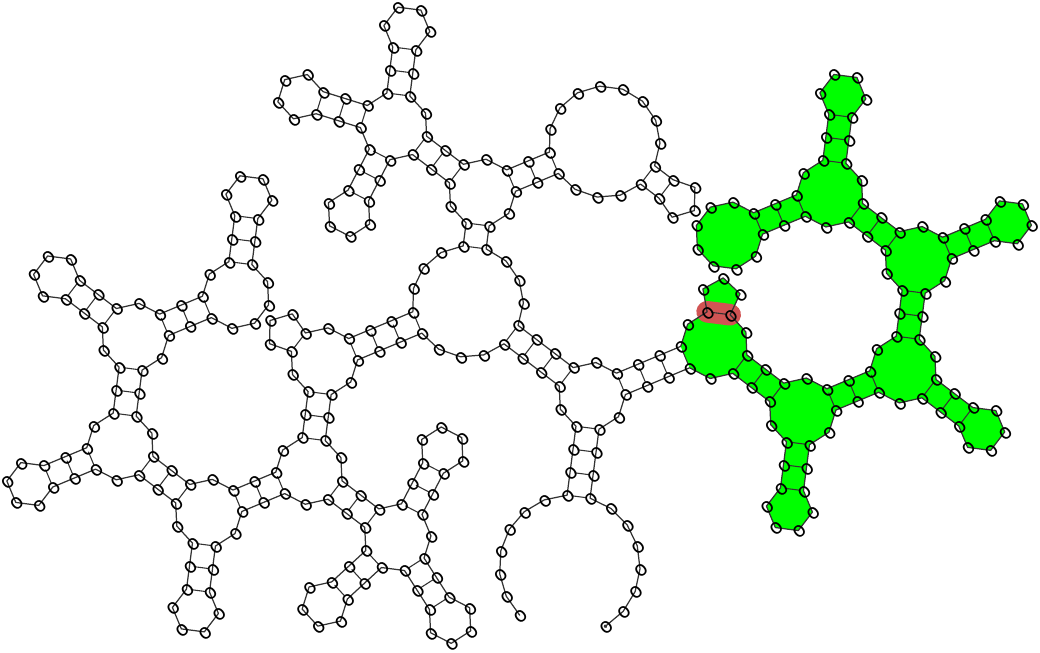}}    \\  \hline
        87 & Cat's Toy 2 & 97 & 1 & alg 3 & 9.68 &\raisebox{-.5\height}{\includegraphics[width=0.09\textwidth, angle=0]{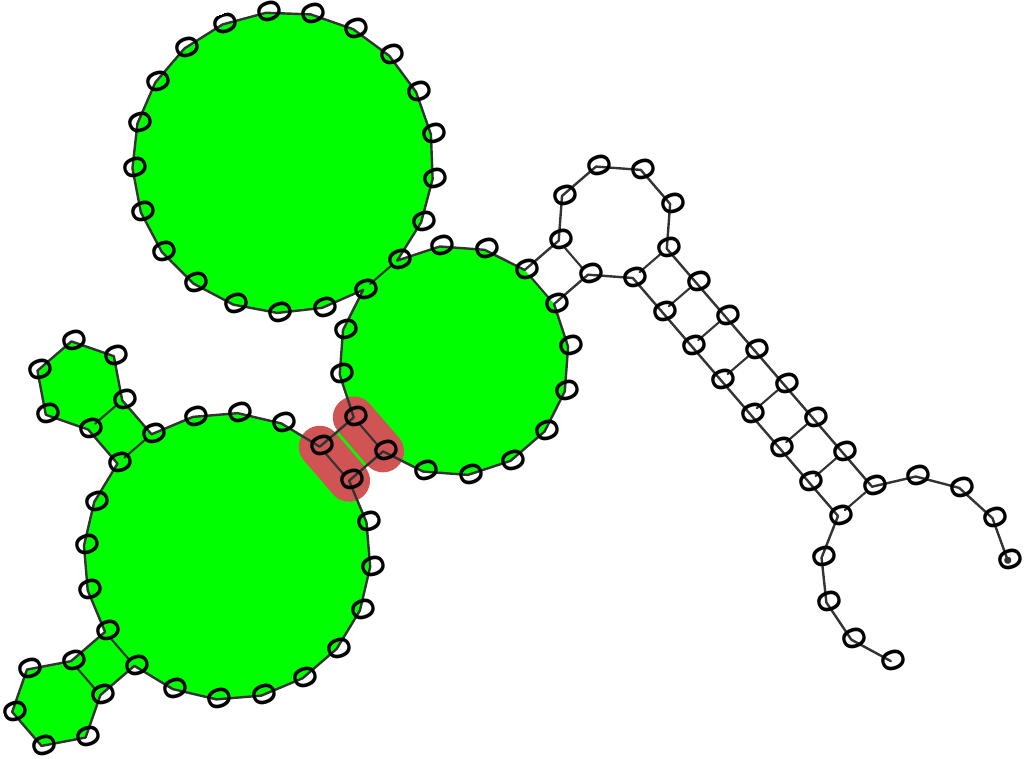}}   \\  \hline
        88 & Zigzag Semicircle & 34 & 9 & alg 2 & 1.51 &\raisebox{-.5\height}{\includegraphics[width=0.09\textwidth, angle=0]{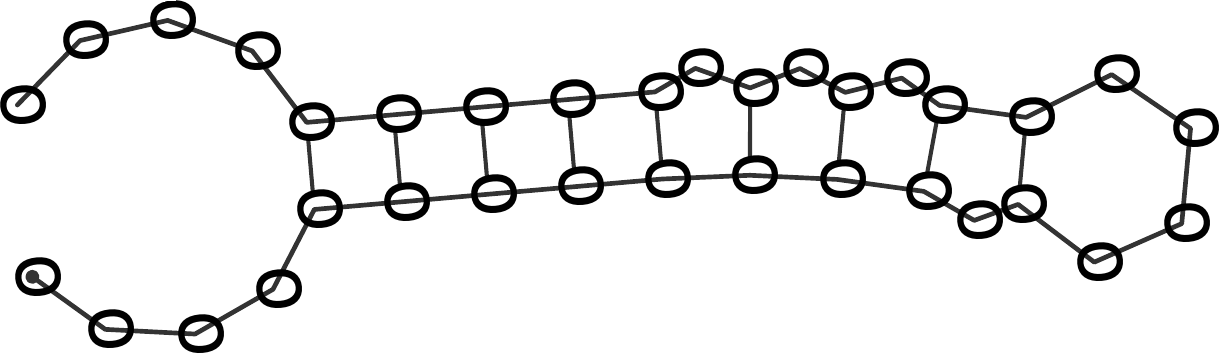}}   \\  \hline
        91 & Thunderbolt & 392 & 1 & alg 3 & 1.04  &\raisebox{-.5\height}{\includegraphics[width=0.23\textwidth, angle=0]{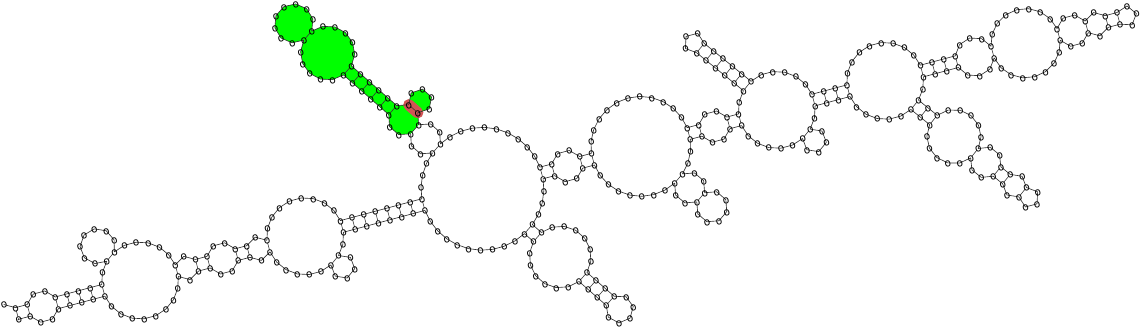}}  \\  \hline
        92 & Mutated chicken feet & 100 & 1 & alg 3 & 223.57 &\raisebox{-.5\height}{\includegraphics[width=0.09\textwidth, angle=0]{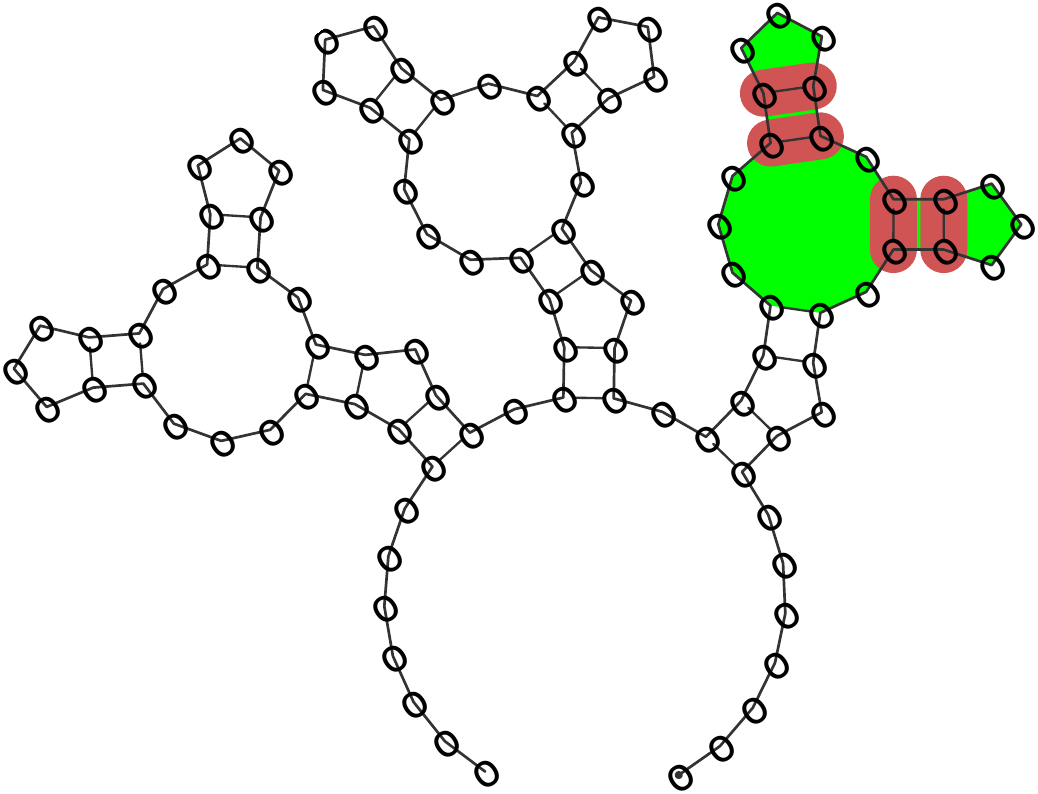}}    \\  \hline
        96 & Cesspool & 358 & 1 & alg 3 & 14.27 &\raisebox{-.5\height}{\includegraphics[width=0.14\textwidth, angle=0]{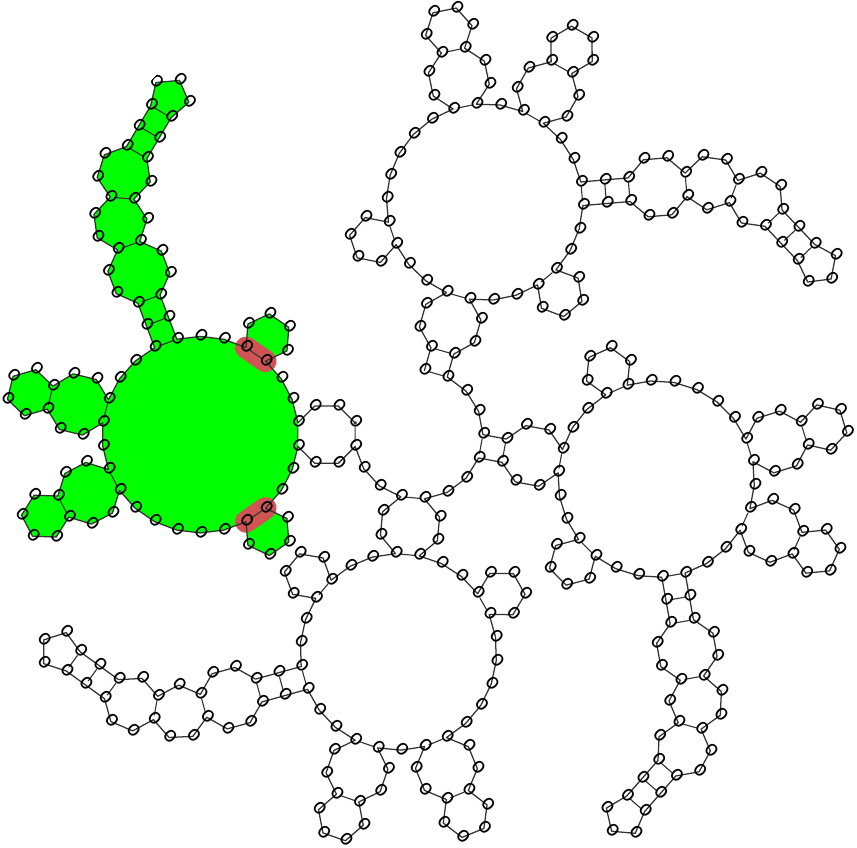}}   \\  \hline
        99 & Shooting Star & 364 & 2 & alg 3 & 7.77 &\raisebox{-.5\height}{\includegraphics[width=0.14\textwidth]{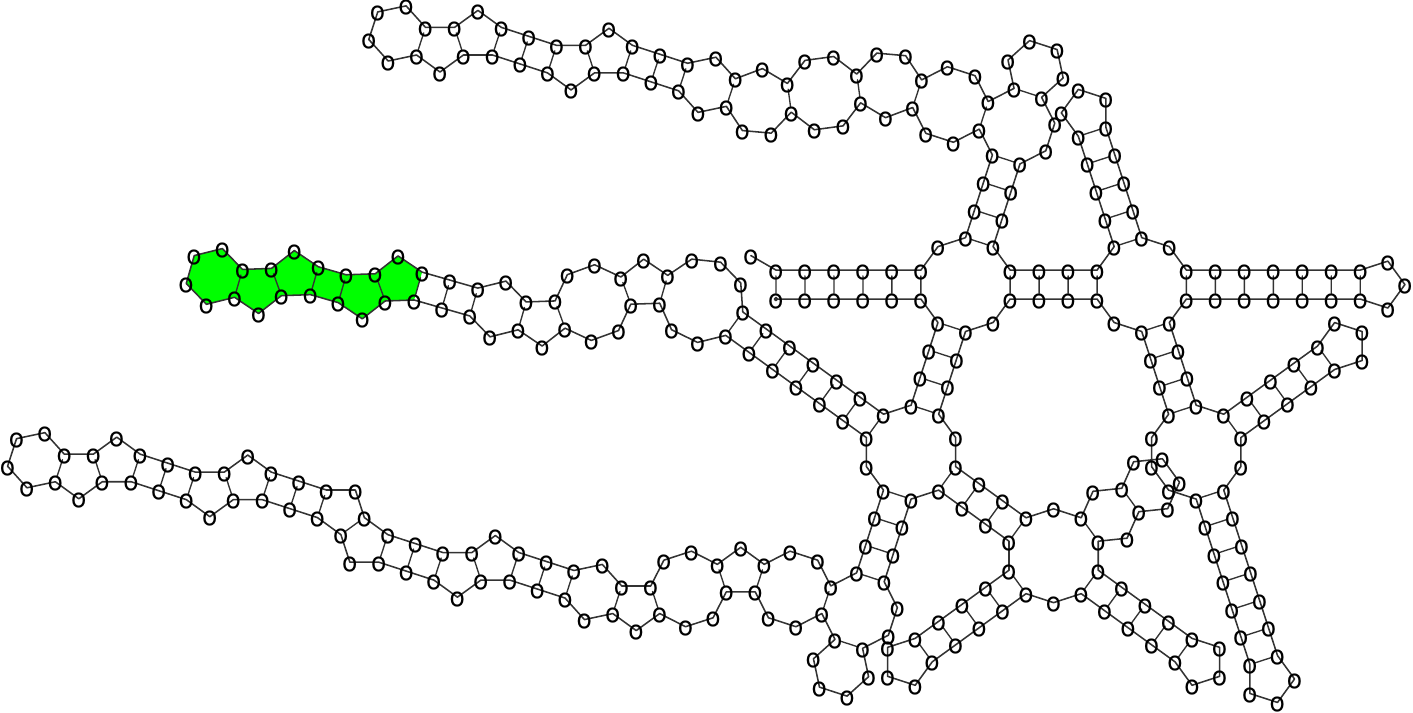}}   \\ \hline
    \end{tabular}
    \label{table:results}
    \end{minipage}
\end{table}
\enlargethispage{\baselineskip}

%% file: conclusion.tex
\section{Conclusions and Future Work}

\label{sec:conclusion}

Following the core idea of proof by construction, we propose three efficient and explainable algorithms (RIGENDE) for \textbf{proving undesignability}  in the context of RNA design with the nearest neighbor model. 
Theoretically, the theorems we introduced can shed some light on why and how some structures are not designable. The establishment of those concepts such as \textbf{rival structure, designability constraint, and context-constrained undesignability} can be regarded as a milestone for automatically verifying undesignability.
Applied to Eterna100 benchmark, RIGENDE can prove 15 of them are actually undesignable using popular Turner model implemented in ViennaRNA 2.5.1 and LinearFold. Without doubt,  the found rival structures can help humans understand more about RNA folding and RNA design. The main drawbacks of RIGENDE include: 
\begin{enumerate}
	\item The rival structure candidates are crucial for the algorithms to work, and the selection of candidates is dependent on the results of RNA design.
	\item The structure decomposition in Algorithm~\ref{alg:3} only considers a pair-bounded substructure, which may not be able to cover other sophisticated cases.
\end{enumerate}

In the future, we would address those drawbacks. We will not only prove more cases of undesignability on structure level but also examine the undesignability on the level of structure motifs.
\begin{enumerate}
	\item Design better ways to find rival structure candidates, such as devising approaches that eliminate the necessity of relying on external RNA design methods.
	\item Decompose structures according to the topology of loops instead of splitting the entire structure into two via a base pair.
	\item Experiment with more puzzles and RNA design settings to search for  more general regularities of \mbox{designability} and undesignability.
\end{enumerate}

%% file: supp.tex

\setcounter{figure}{0}
\renewcommand{\thefigure}{SI\,\arabic{figure}} 
\setcounter{table}{0}
\renewcommand{\thetable}{SI\,\arabic{table}}
\setcounter{page}{1}

\setcounter{section}{0}
\renewcommand\thesection{\Alph{section}}
\setcounter{subsection}{0}
\renewcommand\thesubsection{\Alph{section}.\arabic{subsection}}

\section{Projection and Intersection}\label{sec:op}
\SetKwInOut{KwIn}{Input}
\SetKwInOut{KwOut}{Output}
\SetKwInOut{KwPara}{Parameter}
\SetKwComment{Comment}{/* }{ */}
\setcounter{algocf}{3}


\begin{algorithm}[H]
\NoCaptionOfAlgo
\caption{Projection $\hat{\vecx} = \vecx \proj I$}\label{alg:proj}
\KwIn{$\vecx, I$\Comment*{$I = [i_1, i_2, \ldots, i_n]$ contains the critical positions}}
\KwOut{$\hat{\vecx}$\Comment*{$\hat{\vecx}$ is $\vecx$ projected onto $I$}} 
$\hat{\vecx} \gets \text{map}()$\;  

\For{$i$ in  $ I $}
{
    {$\hat{\vecx}[i] \gets \vecx_i$}
}
\Return $\hat{\vecx}$\
\end{algorithm}


\SetKwInOut{KwIn}{Input}
\SetKwInOut{KwOut}{Output}
\SetKwInOut{KwPara}{Parameter}
\SetKwComment{Comment}{/* }{ */}
\begin{algorithm}
\NoCaptionOfAlgo
\caption{Contraint Intersection $C'_1, C'_2 = \text{Intersection}(C_1, C_2)$}\label{alg:contract}
\KwIn{$C_1, C_2$\Comment*{Constraints from Algorithm 1}}
\KwOut{$C'_1,C'_2$\Comment*{New constraints after removing contradictions}} 

$(I_1, \setxi_1) \gets C_1$;\\
$(I_2, \setxi_2) \gets C_2$;\\
$I' \gets I_1 \cap I_2$;\\

\If{$I' = \emptyset$}{
    $C'_1 \gets (I_1,~\setxi_1)$\;
    $C'_2 \gets (I_2,~\setxi_2)$\;
    \Return $C'_1, C'_2$;
}

$\setxi'_1 \gets \{ \hat{\vecx} \proj I' \mid \hat{\vecx} \in \setxi_1 \}$;\\
$\setxi'_2 \gets \{ \hat{\vecx} \proj I' \mid \hat{\vecx} \in \setxi_2 \}$;

\For{$\hat{\vecx} \in \setxi_1$} {
    \lIf(\tcc*[f]{$\hat{\vecx} \in \setxi_1$ cannot be satisfied with $\setxi_2$}){$\hat{\vecx} \proj I' \notin \setxi'_2$}{$\setxi_1 \gets \setxi_1 \setminus \{\hat{\vecx}\}$}
}

\For{$\hat{\vecx} \in \setxi_2$} {
    \lIf(\tcc*[f]{$\hat{\vecx} \in \setxi_2$ cannot be satisfied with $\setxi_1$}){$\hat{\vecx} \proj I' \notin \setxi'_1$}{$\setxi_2 \gets \setxi_2 \setminus \{\hat{\vecx}\}$}
}
$C'_1 \gets (I_1,~\setxi_1)$\;
$C'_2 \gets (I_2,~\setxi_2)$\;

\Return $C'_1,C'_2$;\ 
\end{algorithm}

\vspace{-1cm}
\section{Energy Scoring Functions}\label{sec:turner}
\subsection{Hairpin}
For the list of special hairpins, see Table~\ref{tab:sp_hairpin}. Set $\text{length} = j - i - 1$.
\begin{equation*}
    \DG(\vecx, H\angle{(i, j)}) = \begin{cases}
        \DG_{\text{special\_hairpin}}(\vecx) &\hspace{-3.7cm}\text{if } (\vecx, H\angle{(i, j)}) \text{ is special hairpin;} \\
        \left (
            \begin{aligned}
                &\DG_{\text{hairpin\_mismatch}}((\vecx_i, \vecx_j), \vecx_{i+1}, \vecx_{j-1}) \\
                &+ \begin{cases}
                    \DG_{\text{hairpin}}(\text{length} ) & \text{if } 0 \leq \text{length} \leq 30 \\
                    \DG_{\text{hairpin}}(30) + \DG_{\text{extend}} \cdot \ln((\text{length}) / 30) & \text{if } \text{length} > 30
                \end{cases}
            \end{aligned}
        \right ) &\text{otherwise.}
    \end{cases}
\end{equation*}

\subsection{Special Hairpin}\label{subsec:sp}
See Table~\ref{tab:sp_hairpin}.
\begin{table}[]
    \centering
    \caption{List of special hairpins in the Turner energy model.}
    \smallskip
    \begin{tabular}[t]{c|c}
        $\vecx$     & $\DG_{\text{special\_hairpin}}(\vecx, \verb|(...)|)$ \\\hline
        \verb|CAACG|    & 6.8                               \\
        \verb|GUUAC|    & 6.9                               \\
        \hline    
    \end{tabular} 
    \begin{tabular}[t]{c|c}
        $\vecx$     & $\DG_{\text{special\_hairpin}}(\vecx, \verb|(....)|)$ \\\hline
        \verb|CAACGG|   & 5.5                               \\
        \verb|CCAAGG|   & 3.3                               \\
        \verb|CCACGG|   & 3.7                               \\
        \verb|CCCAGG|   & 3.4                               \\
        \verb|CCGAGG|   & 3.5                               \\
        \verb|CCGCGG|   & 3.6                               \\
        \verb|CCUAGG|   & 3.7                               \\
        \verb|CCUCGG|   & 2.5                               \\
         \verb|CUAAGG|   & 3.6                               \\
        \verb|CUACGG|   & 2.8                               \\
        \verb|CUCAGG|   & 3.7                               \\
        \verb|CUCCGG|   & 2.7                               \\
        \verb|CUGCGG|   & 2.8                               \\
        \verb|CUUAGG|   & 3.5                               \\
        \verb|CUUCGG|   & 3.7                               \\
        \hline
    \end{tabular}
    \begin{tabular}[t]{c|c}
        $\vecx$     & $\DG_{\text{special\_hairpin}}(\vecx, \verb|(......)|)$ \\\hline
        \verb|ACAGUACU| & 2.8                               \\
        \verb|ACAGUGCU| & 2.9                               \\
        \verb|ACAGUGAU| & 3.6                               \\
        \verb|ACAGUUCU| & 1.8                               \\
        \hline
    \end{tabular}
    \label{tab:sp_hairpin}
\end{table}
\vspace{-0.5cm}
\subsection{Stack}
\vspace{-1cm}
\begin{equation*}
    \DG(\vecx, S\angle{(i, j), (k, l)}) = \DG_{\text{stack}}((\vecx_i, \vecx_j), (\vecx_k, \vecx_l))
\end{equation*}
\vspace{-0.5cm}
\subsection{Bulge}
Set $\text{length} = k - i + j - l - 2$.\\
\begin{multline*}
    \DG(\vecx, B\angle{(i, j), (k, l)}) = \begin{cases}
        \DG_{\text{bulge}}(\text{length}) & \text{if } \text{length} \leq 30\\ 
        \DG_{\text{bulge}}(30) + \DG_{\text{extend}} \cdot \ln(\text{length} / 30) & \text{if } \text{length} > 30
    \end{cases}\\
    + \begin{cases}
        \DG_{\text{stack}}((\vecx_i, \vecx_j), (\vecx_k, \vecx_l)) &\text{if } \text{length} = 1\\
        \left (
            \begin{cases}
                \DG_{\text{AU/GU}} &\text{if } \vecx_i = \text{U} \text{ or } \vecx_j = \text{U}\\
                0 &\text{otherwise}
            \end{cases} +
            \begin{cases}
                \DG_{\text{AU/GU}} &\text{if } \vecx_k = \text{U} \text{ or } \vecx_l = \text{U}\\
                0 &\text{otherwise}
            \end{cases}
        \right ) &\text{otherwise}
    \end{cases}
\end{multline*}

\subsection{Internal Loop}
For $1 \times 1$, $1 \times 2$, $2 \times 1$, $2 \times 2$, $2 \times 3$, and $1 \times n$ special internal loops, see Supplementary Section \ref{sec:sp_internal}. Set $\text{length} = (k - i) + (j - l) - 2$.
\begin{multline*}
    \DG(\vecx, I\angle{(i, j), (k, l)}) = \DG_{\text{internal\_mismatch}}((\vecx_i, \vecx_j), \vecx_{i+1}, \vecx_{j-1}) \\
    + \DG_{\text{internal\_mismatch}}((\vecx_k, \vecx_l), \vecx_{k-1}, \vecx_{l+1})\\
    + \min(n_{37}^{\max}, n_{37} \cdot \left | (k - i) - (j - l) \right |)\\
    + \begin{cases}
        \DG_{\text{internal}}(\text{length}) & \text{if } \text{length} \leq 30\\ 
        \DG_{\text{internal}}(30) + \DG_{\text{extend}} \cdot \ln([\text{length}] / 30) & \text{if } \text{length} > 30
    \end{cases}\\
\end{multline*}

\subsection{Special Internal Loop}\label{sec:sp_internal}
Set $\text{length} = k - i + j - l - 1$,
\begin{align*}
    &\DG_{\text{special\_internal}}(\vecx, I\angle{(i, j), (k, l)}) = \\
    &\begin{cases}
        \DG_{\text{internal\_11}}((\vecx_i, \vecx_j), (\vecx_k, \vecx_l), \vecx_{i+1}, \vecx_{j-1})& \text{if } 1 \times 1\\
        \DG_{\text{internal\_12}}((\vecx_i, \vecx_j), (\vecx_k, \vecx_l), \vecx_{i+1}, \vecx_{k-1}, \vecx_{j-1})& \text{if } 1 \times 2\\
        \DG_{\text{internal\_21}}((\vecx_i, \vecx_j), (\vecx_k, \vecx_l), \vecx_{i+1}, \vecx_{l+1}, \vecx_{j-1})& \text{if } 2 \times 1\\
        \DG_{\text{internal\_22}}((\vecx_i, \vecx_j), (\vecx_k, \vecx_l), \vecx_{i+1}, \vecx_{k-1} \vecx_{l+1}, \vecx_{j-1})& \text{if } 2 \times 2\\
        \left (
        \begin{aligned}
            &\DG_{\text{internal}}(5) + n_{37}\\
            &+ \DG_{\text{internal\_23\_mismatch}}((\vecx_i, \vecx_j), \vecx_{i+1}, \vecx_{j-1})\\
            &+ \DG_{\text{internal\_23\_mismatch}}((\vecx_k, \vecx_l), \vecx_{k-1}, \vecx_{l+1})
        \end{aligned} \right ) & \text{if } 2 \times 3\\
        \left (
        \begin{aligned}
            &\DG_{\text{internal\_1n\_mismatch}}((\vecx_i, \vecx_j), \vecx_{i+1}, \vecx_{j-1}) \\
            &+ \DG_{\text{internal\_1n\_mismatch}}((\vecx_k, \vecx_l), \vecx_{k-1}, \vecx_{l+1})\\
            &+ \min(n_{37}^{\max}, n_{37} \cdot \left | (k - i) - (j - l) \right |)\\
            &+ \begin{cases}
                \DG_{\text{internal}}(\text{length}+1) & \text{if } \text{length}+1 \leq 30\\ 
                \DG_{\text{internal}}(30) + \DG_{\text{extend}} \cdot \ln([\text{length} + 1] / 30) & \text{if } (k - i) + (j - l) - 2 > 30
            \end{cases}
        \end{aligned} \right ) & \text{if } 1 \times n\\
    \end{cases}
\end{align*}

\subsection{Multiloop}
Let $m = (i, j), (i_1, j_1), (i_2, j_2), \ldots, (i_k, j_k)$
\vspace{-0.5cm}
\begin{equation*}
    \DG(\vecx, M\angle{m}) = \DG_{\text{closing\_pair}}(i, j) + \sum_{l=1}^k \DG_{\text{enclosed\_pair}} (\vecx, i_l, j_l)
\end{equation*}
\vspace{-1.5cm}
\begin{figure}
    \hspace{-0.0cm}
    \begin{minipage}{0.45\textwidth}
        \begin{align*}
            \DG_{\text{closing\_pair}}(i, j) &= \DG_{\text{closing}}\\
             &+ \DG_{\text{multi\_mismatch}}((\vecx_i, \vecx_j), \vecx_{i+1}, \vecx_{j-1})\\
             &+ \DG_{\text{multi\_internal}} \\
             &+ \begin{cases}
                \DG_{\text{AU/GU}} &\text{if } \vecx_i = \text{U} \text{ or } \vecx_j = \text{U}\\
                0 &\text{otherwise}
            \end{cases}
        \end{align*}
    \end{minipage}
    \hfill
    \hspace{0.0cm}
    \begin{minipage}{0.45\textwidth}
        \begin{align*}
            \DG_{\text{enclosed\_pair}}(i, j) &= \DG_{\text{multi\_mismatch}}((\vecx_i, \vecx_j), \vecx_{i-1}, \vecx_{j+1})\\
             &+ \DG_{\text{multi\_internal}} \\
             &+ \begin{cases}
                \DG_{\text{AU/GU}} &\text{if } \vecx_i = \text{U} \text{ or } \vecx_j = \text{U}\\
                0 &\text{otherwise}
            \end{cases}
        \end{align*}
    \end{minipage}
    \caption{Side by side alignment of equations.}
\end{figure}
\vspace{-1.0cm}
\subsection{External Loop}
\begin{align*}
    \DG(\vecx, E\angle{(5', 3'), (i, j)}) = &\DG_{\text{external\_mismatch}}((\vecx_i, \vecx_j), \vecx_{i-1}, \vecx_{j+1}) \\
    &+ \DG_{\text{dangle\_5'}}((\vecx_i, \vecx_j), \vecx_{i-1})
    + \DG_{\text{dangle\_3'}}((\vecx_i, \vecx_j), \vecx_{j+1})\\
    &+ \begin{cases}
        \DG_{\text{AU/GU}} &\text{if } \vecx_i = \text{U} \text{ or } \vecx_j = \text{U}\\
        0 &\text{otherwise}
    \end{cases}\\
\end{align*}

\vspace{-1cm}
\section{Designable Case}\label{sec:dcase}
The puzzle ``Short String 4" and a designed sequence are shown in Fig.~\ref{fig:shortstr4}.
\begin{figure}
\centering
  \centering
  \scalebox{0.8}{%
  \begin{tabular}{rl}
 $\ystar$: &	\texttt{......((.((..((..(...(.((.((....))))..)...(.((..((....))..))..)...).))..)))).....................}\\
 $\vecx$: &	\texttt{CAAGAAGCGCGGACUGACAGAGAGGAUCGAAAGACUGACAAAGAGAGGGCGAAAGCAAUUGACAAAGAGGGACGGUAAAAAAAAAAAAAAAAAAGAA}\\
 \end{tabular}%
 }
  \caption{The puzzle ``Short String 4" ($\ystar$) and a designed sequence ($\vecx$)}
  \label{fig:shortstr4}
\end{figure}
\vspace{-1cm}